\documentclass[12pt]{article}
\usepackage{feynmf}
\def\tf#1#2{{\textstyle{#1 \over #2}}}

\def\overleftrightarrow#1{\vbox{\ialign{##\crcr
     $\leftrightarrow$\crcr\noalign{\kern-0pt\nointerlineskip}
     $\hfil\displaystyle{#1}\hfil$\crcr}}}

\def\sqr#1#2{{\vcenter{\vbox{\hrule height.#2pt
         \hbox{\vrule width.#2pt height#1pt \kern#1pt
            \vrule width.#2pt}
         \hrule height.#2pt}}}}
\def\square{\mathop{\mathchoice\sqr56\sqr56\sqr{3.75}4\sqr34\,}\nolimits}
\begin{document}
\baselineskip=15.5pt
\pagestyle{plain}
\setcounter{page}{1}
\renewcommand{\theequation}{\thesection.\arabic{equation}}
\setlength{\unitlength}{1mm}
\begin{fmffile}{fmfdefect}
%--------+---------+---------+---------+---------+---------+---------+
%Title page
\begin{titlepage}
\bigskip
\rightline{CALT-68-2359}
\rightline{CITUSC/01-041}
\rightline{NSF-ITP-01-172}
\rightline{MIT-CTP-3212}
\rightline{hep-th/0111135}
\bigskip\bigskip\bigskip\bigskip
\centerline{\Large \bf Holography and Defect Conformal Field Theories}
\bigskip\bigskip\bigskip

\centerline{\large Oliver DeWolfe$^a$, Daniel Z.\ Freedman$^{a,b}$ and
Hirosi Ooguri$^{a,c}$}
\bigskip
\centerline{\em $^a$Institute for Theoretical Physics, UCSB, Santa
Barbara, CA 93106}
\smallskip
\centerline{\em $^b$ Department of Mathematics and Center for
Theoretical Physics,} \centerline{\em Massachusetts Institute of
Technology, Cambridge, MA, 02139}
\smallskip
\centerline{\em $^c$California Institute of Technology 452-48,
Pasadena, CA, 91125}
\medskip
\bigskip\bigskip

%ABSTRACT

\begin{abstract}
We develop both the gravity and field theory sides of the
Karch-Randall conjecture that the near-horizon description of a
certain D5-D3 brane configuration in string theory, realized as $AdS_5
\times S^5$ bisected by an $AdS_4 \times S^2$ ``brane'', is dual to
${\cal N} =4$ Super Yang-Mills theory in ${\bf R}^4$ coupled to an
${\bf R}^3$ defect.  We propose a complete Lagrangian for the field
theory dual, a novel ``defect superconformal field theory'' wherein a
subset of the fields of ${\cal N} =4$ SYM interacts with a $d=3~SU(N)$
fundamental hypermultiplet on the defect preserving conformal
invariance and 8 supercharges.  The Kaluza-Klein reduction of wrapped
D5 modes on $AdS_4 \times S^2$ leads to towers of short
representations of $OSp(4|4)$, and we construct the map to a set of
dual gauge-invariant defect operators ${\cal O}_3$ possessing integer
conformal dimensions.  Gravity calculations of $ \langle {\cal O}_4
\rangle$ and $\langle {\cal O}_4 \, {\cal O}_3 \rangle$ are given.
Spacetime and $N$-dependence matches expectations from dCFT, while
the behavior as functions of $\lambda=g^2N$ at strong and weak
coupling is generically different.  We comment on a class of
correlators for which a non-renormalization theorem may still exist.
Partial evidence for the conformality of the quantum theory is given,
including a complete argument for the special case of a $U(1)$ gauge
group.  Some weak coupling arguments which illuminate the duality are
presented.

\medskip
\noindent
\end{abstract}
\end{titlepage}

%--------+---------+---------+---------+---------+---------+---------+
%Body

\section{Introduction}

The study of the $AdS$/CFT correspondence \cite{Maldacena, GKP,
Witten} (for a review, see \cite{MAGOO}) has taught us much about both
the behavior of field theories and the nature of string theory.
Accordingly, generalizations of the correspondence with additional
structure added to both sides are inherently quite interesting, and
potentially have much more to teach us about field theory dynamics,
the nature of string theory and how holography relates them.

It is well-known that spatial defects may be introduced into conformal
field theories, reducing the total symmetry but preserving conformal
invariance \cite{Cardy, McO}.  Whether one can obtain holographic
duals of such ``defect conformal field theories'' (dCFTs) is a
fascinating question.  A potential gravity dual was proposed by Karch
and Randall \cite{KR1}, who studied curved branes in anti-de Sitter
space in an effort to ``locally localize'' gravity \cite{KR2}.

In their investigation, Karch and Randall noticed that an $AdS_4$
brane inside $AdS_5$ could be naturally realized in string theory
using a certain D3-brane/D5-brane system.  The near-horizon limit of
the $N$ D3-branes produces an $AdS_5 \times S^5$ background in which
the D5-branes occupy an $AdS_4 \times S^2$ submanifold.  Karch and
Randall speculated that the $AdS$/CFT correspondence would ``act
twice'' in this system, meaning that in addition to the closed strings
propagating throughout space providing a holographic description of a
field theory on the boundary of $AdS_5$ as usual, the fluctuations on
the $AdS_4$ brane should be dual to additional physics confined to the
boundary of the $AdS_4$. Hence, the dual field theory
contains not only the usual $d=(3+1)~{\cal N}=4$ Super-Yang Mills
theory, but also new fields and couplings living on a
$(2+1)$-dimensional defect, obtained from the low-energy limit of the
3-5 open strings interacting with the 3-3 strings of the original
brane setup.

We study the case of a single D5-brane, whose backreaction on the
near-horizon geometry can be neglected in the 't Hooft limit,
allowing it to be treated as a probe hosting open strings.  The
resulting dual field theory consists of $SU(N)$ ${\cal N}=4$ SYM in
{\bf R}$^4$, with a subset of these ambient modes interacting in a
fashion we will determine with a single fundamental hypermultiplet on
the {\bf R}$^3$ defect.  The resulting theory has half the
supersymmetry of the ambient theory, but intriguingly, must preserve
$SO(3,2)$ conformal symmetry in order to match the unbroken anti-de
Sitter isometries on the gravity side.  As a result the Karch-Randall
system is an ideal candidate for the holographic description of a
dCFT.  We will construct the field theory explicitly as a novel defect
superconformal theory with an exact Lagrangian description.

The reduced symmetries of codimension-one dCFTs admit interesting
structures such as one-point functions for the usual operators in the
ambient space, two-point functions for ambient operators with
different conformal dimensions, and mixed two-point functions between
these and operators localized on the defect; the functional forms of
such correlators are significantly constrained \cite{Cardy,McO}.  On
the supergravity side, we employ holography to calculate such novel
correlation functions from Witten diagrams involving integrals over
the $AdS_4$ submanifold, and we reproduce the space-time forms
required by defect conformal symmetry.

We consider the expansion of the D5-brane action through quadratic
order in fluctuations about the $AdS_4 \times S^2$ probe
configuation. We perform a Kaluza-Klein reduction of quadratic terms
in bosonic open string fields $(\psi)$ and find a set of modes of
integer mass and scale dimension. The lowest mode of the D5-brane
gauge field on $AdS_4$ is dual to the current of a global $U(1)_B$
symmetry in the field theory.  As expected all modes can be organized
in short representations of the superalgebra $OSp(4|4)$ associated
with supersymmetry in $AdS_4$. Other terms in the fluctuation action
involve closed string fields $(\phi)$, specifically terms of order
$\phi, \phi\psi$, and $\phi^2$. These are interpreted as interactions
which determine the novel correlators discussed above. We also obtain
the leading power of $N$ and the 't Hooft coupling $\lambda$ for the
D5-brane contribution to all correlation functions, a strong-coupling
prediction.

We then turn to the dual dSCFT.  Using gauge invariance, supersymmetry
and R-symmetry, we construct the field theory Lagrangian.  This
involves augmenting the usual ${\cal N}=4$ Super Yang-Mills in four
dimensions with dynamics on the defect.  The fundamental defect
hypermultiplet couples canonically to the restriction of the 4D gauge
field to the hypersurface; we use the ``superspace boundary''
technique \cite{Hori} to derive a defect action preserving eight
supercharges.  We construct the action in ${\cal N}=1$ superspace, but
demonstrate that it is fully ${\cal N}=4$ supersymmetric by
identifying the $SU(2)_V \times SU(2)_H$ R-symmetry.  The symmetries
rule out any additional marginal interactions, preserving the 4D gauge
coupling $g$ as the only dimensionless parameter, as well as
forbidding mass terms, leaving the theory classically
conformal-invariant.  Interestingly, the bulk fields participating in
the defect interaction involve not just half the scalars, but the
normal derivatives of the other half.  The bosonic parts of related
(non-conformal) supersymmetric defect actions derived from intersecting branes
appeared in \cite{Sav, KS}.

We also match the bosonic modes of the D5-brane on the gravity side to
dual field theory operators.  The multiplets are short, so conformal
dimensions should be protected in the usual way.  There is a unique
candidate for the chiral primary operator of the lowest multiplet, and
we use supersymmetry to fill out the rest of this multiplet, matching
the modes to fluctuations on the gravity side.  We also discuss the
operator structure of higher multiplets. Weak coupling calculations
help to determine which operators have protected scale dimensions.

Finally, we discuss the perturbative dynamics of the field theory.  We
argue that for a certain class of ``pinned'' correlators, there are no
divergences other than wavefunction renormalization of the defect
fields.  This is sufficient to demonstrate quantum conformal
invariance for gauge group $U(1)$. For gauge group $SU(N)$ non-pinned
correlators must be considered as well, and we have not yet studied
these.  Hence the question of quantum conformal invariance remains
open.  We also discuss the field theory computation of various one-
and two-point functions, and compare to gravity.  We find that
although the powers of $N$ match perfectly, the powers of the 't Hooft
parameter do not.  Hence, unlike the ${\cal N}=4$ case, the simplest
correlators of this theory do not obey a nonrenormalization theorem.
We do describe a class of correlators independent of $\lambda$ at
leading order, for which a non-renormalization theorem is not ruled
out.  We conclude with a discussion of directions for future research.

One can consider analogous models in other dimensions.
Defect conformal field theories in two dimensions are
studied in \cite{BdBDO}. Some of them have holographic
duals in $AdS_3$, such as the $AdS_2$ branes inside
$AdS_3$ with the NS-NS flux studied in \cite{Bachas}.
In these cases, one may be able to study the
correspondence beyond the supergravity approximation.

Sections \ref{FieldSec} and \ref{PerturbSec} on the construction and
analysis of the field theory are largely independent of holography and
can be read separately.

\setcounter{equation}{0}
\section{Description of the System}

\subsection{Brane construction}

The system of partially-overlapping 3-branes and 5-branes preserving 8
supercharges has been known for some time, and was extensively studied
in \cite{HW} as a way to engineer 3-dimensional ${\cal N}=4$ field
theories on branes.  In contrast, we consider systems which have
infinite D3-branes, and hence have four-dimensional (as well as
three-dimensional) dynamics.

We choose coordinates as follows.  The $N$ D3-branes fill the 0126
directions, while the D5-brane spans 012345; all the branes sit at the
origin of the transverse coordinates.  In the absence of the D5-brane,
the system has 16 unbroken supercharges, an $SO(3,1)$ Lorentz symmetry
acting on $(x_0, x_1, x_2, x_6)$ and an additional $SO(6) \sim SU(4)$
acting on $(x_3, x_4, x_5, x_7, x_8, x_9)$. The D3-D5 background
preserves 8 supersymmetries, reduces $SO(3,1)$ to $SO(2,1)$ on $(x_0,
x_1, x_2)$, and breaks $SO(6)$ to $SO(3) \times SO(3) \sim SU(2)_H
\times SU(2)_V$ acting on $(x_3, x_4, x_5)$ and $(x_7, x_8, x_9)$,
respectively.

Four kinds of strings exist in this system.  As usual, closed
strings propagate in the bulk, giving rise to the fields of IIB SUGRA
as well as all the excited modes. Also, 3-3 and 5-5 open strings
lead to sixteen-supercharge vector multiplets on the D3-brane and
D5-brane, respectively; these each split into a vector multiplet and a
hypermultiplet under the preserved eight supercharges.  Finally, 3-5
strings localized on the (2+1)-dimensional intersection of the branes
lead to a three-dimensional hypermultiplet, charged as a bifundamental
under the gauge group of each brane.  

\subsection{Near-horizon Limit}

We remind the reader of the familiar facts of the original $AdS$/CFT
procedure of Maldacena \cite{Maldacena}.  Consider a stack of $N$ parallel
D3-branes with $g_s \rightarrow 0$, $N \rightarrow \infty$ with $g_s
N$ fixed.  This system may be examined either for $g_s N \ll 1$, in
which case the appropriate description is provided by open strings
propagating on flat branes, or for $g_s N \gg 1$, in which case the
appropriate description is a black three-brane solution of Type IIB
supergravity.  By sending $l_s \rightarrow 0$ with the energies of
fluctuations fixed, one is left in the former case with the
renormalizable field theory of the massless open string states, namely
4D ${\cal N}=4$ Super-Yang Mills, and with closed strings propagating
in the $AdS_5 \times S^5$ near-horizon geometry of the black brane in
the latter case.

Thus, the two kinds of string modes in the original brane set-up, open
and closed, have been segregated from one another, yet are found to
describe the same physics in the field theory/near-horizon limit.
Each description is useful in a different region of parameter space.
Additionally, the symmetry groups enlarge on both sides in the limit,
as the field theory is exactly (super)conformal, while $AdS$
isometries appear on the gravity side; the 4D conformal and 5D anti-de
Sitter supergroups are algebraically identical, and are denoted
$SU(2,2|4)$.  This group also contains the $SO(6) \sim SU(4)$ of the
original brane setup, which is an R-symmetry in the field theory and
the isometry group of $S^5$ in the dual.

The system we study is richer, but displays similar behavior.  Again
we take $g_s \ll 1$, $N \gg 1$ with $\lambda \equiv g_s N$ fixed.  For
the case $g_s N \ll 1$, the appropriate description of the branes are
as flat hypersurfaces.  We take the limit $l_s \rightarrow 0$ with the
energies of D3-brane fluctuations fixed.  This decouples the modes of
the heavy D5-branes, as in \cite{HW}, and leads to the
(3+1)-dimensional field theory described by ${\cal N}=4$ SYM
throughout most of the space, but with a (2+1)-dimensional defect
containing a localized, interacting fundamental hypermultiplet.

For $g_s N \gg 1$, on the other hand, the appropriate description of
the D3-branes is a black three-brane.  However, we still have $g_s \ll
1$, and hence a single D5-brane should still be described as a
hypersurface with propagating open strings.  Taking the $l_s
\rightarrow 0$ limit here leads to the usual $AdS_5 \times S^5$
near-horizon geometry of D3-branes with an embedded ``probe''
D5-brane.\footnote{Locally localizing gravity the D3/D5 system
requires $M$ D5-branes with $g_sM \gg 1$, a different regime from our
case \cite{KR1}.  Other studies of $AdS_4/AdS_5$ setups with strong
backreaction include \cite{KKR, Poratti}.} Once again the stringy
modes of the brane set-up have been segregated into two sets, one for
each limit of $g_s N$: the closed strings and open 5-5 strings
describe the gravity side, while the low-energy limit of the 3-3 and
3-5 open strings produces the field theory.  Once again, the
expectation is that the two systems are holographic duals of one
another.

We may readily see that the D5-brane lives on an $AdS_4 \times S^2$
submanifold of $AdS_5 \times S^5$, as follows.  In the near-horizon
geometry of the D3-branes, the useful coordinates are $\vec{y} \equiv
(x_0, x_1, x_2)$, $x \equiv x_6$, and the radial coordinate $v$ and
the angles $\Omega_5 = (\psi, \theta, \varphi, \chi, \varsigma)$ defined by
\begin{eqnarray}
\label{DefineCoords}
x_3 = v \cos\psi \sin\theta \cos\varphi \,, \quad \quad 
x_4 &=& v \cos\psi \sin\theta \sin\varphi \,, \quad \quad
x_5 = v \cos\psi \cos\theta \,, \\
x_7 = v \sin\psi \sin\chi \cos\varsigma \,, \quad \quad
x_8 &=& v \sin\psi \sin\chi \sin\varsigma \,, \quad \quad
x_9 = v \sin\psi \cos\chi \,. \nonumber
\end{eqnarray}
The metric for the near-horizon geometry in this coordinate system is
\begin{eqnarray}
\label{Metric}
ds^2 &=& ds^2_{AdS_5} + ds^2_{S^5} \,, \\ ds^2_{AdS_5} &=& L^2 \left( {dv^2 \over v^2} + v^2( dx^2 + d\vec{y}^2) \right)\,, \label{AdSMetric} \\ ds^2_{S^5} &=& L^2 \left(d\psi^2 +
\cos^2 \psi (d \theta^2 + \sin^2 \theta d \varphi^2) + \sin^2 \psi (d
\chi^2 + \sin^2 \chi d \varsigma^2) \right) \,, \label{SphereMetric}
\end{eqnarray}
where as usual $L^4 = 4 \pi \alpha'^2 g_s N$.  The D5-brane sits at $x
= \psi = 0$, filling the $AdS_4$ defined by the coordinates $v$,
$\vec{y}$ and wrapping the $S^2$ parameterized by
$\theta$, $\varphi$.  

The isometry group of the metric (\ref{AdSMetric}),
(\ref{SphereMetric}) preserved by the D5-brane is $SO(3,2) \times
SU(2)_V \times SU(2)_H$.  $SO(3,2)$ acts on $(v,\vec{y})$, while while
$SU(2)_H$ and $SU(2)_V$ rotate $(\theta, \varphi)$ and $(\chi,
\varsigma)$, respectively.  From a field theory viewpoint $SU(2)_V
\times SU(2)_H$ is the unbroken R-symmetry and $SO(3,2)$ is the 3D
conformal group, suggesting that the dual field theory must be exactly
conformal and contain the eight preserved supercharges of the D3-D5
system.  Including the superconformal enehancement to sixteen
supercharges, we expect to find the supergroup $OSp(4|4)$.

\subsection{Correlators in a defect CFT}

The symmetries and the form of correlation functions for CFT$_d$ with
planar boundary have been discussed in the literature, for example in
\cite{Cardy, McO}. Our field theory system, a CFT$_4$ in ${\bf R}^4$
with additional fields on a planar ${\bf R}^3$ defect, shares these
features. We therefore review the most relevant part of this
information, which is mostly taken from \cite{McO}.

In the field theory description we denote points of ${\bf R}^4$ by
$(\vec{y}, x) = x_\mu$ with the defect at $x=0$. The $SO(3,2)$
conformal group of the dCFT is generated by 3-dimensional
translations and Lorentz transformations together with the
4-dimensional inversion, $x_\mu \rightarrow x_\mu/(x_\nu
x^\nu)$. These transformations preserve the defect and act on it as
standard 3-dimensional conformal transformations.

The possible forms of correlation functions for primary scalar
operators ${\cal O}_4$ on the ambient ${\bf R}_4$ and ${\cal O}_3$ on
the defect are restricted by the conformal symmetry. Correlators
involving only ${\cal O}_3$ have the properties expected from standard
3-dimensional conformal invariance, e.g.~the space-time form of two-
and three-point functions is completely determined, while four-point
functions contain an arbitrary function of two ``cross-ratio''
variables.

On the other hand the restriction of the conventional conformal group
$SO(4,2)$ of CFT$_4$ to $SO(3,2)$ leads to new possibilities for
correlators of ${\cal O}_4$ in dCFT.  Let scale dimensions of
operators ${\cal O}_4$ and ${\cal O}_3$ be denoted by $\Delta_4$ and
$\Delta_3$, respectively.  There are non-vanishing one-point functions
$\langle {\cal O}_4 \rangle$, with fully determined space-time
dependence:
\begin{eqnarray}
\label{One}
\langle  {\cal O}_4(x,\vec{y})\rangle &=& {c \over x^{\Delta_4}} \,,
\end{eqnarray}
as well as two-point functions $\langle {\cal O}_4 \, {\cal O}_3
\rangle$ between one ambient and one defect operator, with space-time dependence also fully determined:
\begin{eqnarray}
\langle {\cal O}_4(x,\vec{y}) \, {\cal O}_3(\vec{y}') \rangle
&=& {c' \over {x^{\Delta_4-\Delta_3} \, \eta^{\Delta_3}}} \,, \quad \quad 
\eta \equiv x^2 +(\vec{y}-\vec{y}')^2 \,,
\label{Mixed} 
\end{eqnarray}
and finally there can be non-vanishing two-point functions $\langle
{\cal O}_4 \, {{\cal O}_4}' \rangle$ between ambient operators with
$\Delta_4 \ne {\Delta_4}'$, containing an arbitrary function of one
invariant variable:
\begin{eqnarray}
\langle {\cal O}_4(x,\vec{y}) \, {\cal O}_4(x',\vec{y'})' \rangle
 &=&  {1 \over {x^{\Delta_4}\, {x'}^{\Delta_4'}}} \, f(\xi) \,, \quad \quad
\xi \equiv (x_\mu-{x_\mu}')^2/4xx' \,. 
\label{CorrectedTwo} 
\end{eqnarray}
Our calculations in both weak coupling field theory and the $AdS_5/AdS_4$ 
dual confirm this structure.

On the gravity side the action of the conformal symmetries is best
seen if we transform the radial coordinate $v$ to $z \equiv 1/v$, in
terms of which the $AdS_5$ metric (\ref{AdSMetric}) becomes
conformally flat,
\begin{eqnarray}
\label{zMetric}
ds^2_{AdS_5} &=& {L^2 \over z^2} \left( dz^2 + dx^2 + d\vec{y}^2
\right)\,. 
\end{eqnarray}
The boundary is now at $z=0$. The usual inversion isometry of $AdS_5$
preserves both the boundary and the $AdS_4$ of the D5-brane at $x=0$.
It acts as the standard inversion on this $AdS_4$. Hence the usual
relation between bulk isometries and conformal symmetries on the boundary 
of the usual $AdS$/CFT correspondence extends to the new $AdS_5/AdS_4$
setup.\footnote{The 5D inversion
also preserves the more general Karch-Randall $AdS_4$ surfaces at
$x=Cz$ and acts as the standard inversion on these surfaces.}

\setcounter{equation}{0}
\section{String theory side}
\label{GravitySec}

The bulk degrees of freedom at $g_s \rightarrow 0$, $g_s N$ fixed but
large include both closed string modes, and open string excitations on
the D5-brane.  The former are the massless multiplet of IIB SUGRA
reduced on $AdS_5 \times S^5$, while the latter are a 6D
16-supercharge vector multiplet living on the D5, dimensionally
reduced on $AdS_4 \times S^2$.

With the goal of calculating correlation functions, we are interested
in the fluctuation equations of this system.  The total action is the
sum of the IIB SUGRA action and the Born-Infeld and Wess-Zumino pieces
of the D5-brane action:
\begin{eqnarray}
\label{TotalAction}
S_{tot} = S_{IIB} + S_{BI} + S_{WZ} \,.
\end{eqnarray}
The fluctuation equations for IIB SUGRA reduced on $AdS_5 \times S^5$
were analyzed in \cite{KRvN}, and they have been used extensively to
calculate correlations for gauge-invariant operators ${\cal O}_4$ in
${\cal N}=4$ SYM at large $\lambda$ (for a review and references, see
\cite{MAGOO}).  These results will generically be corrected in our
system due to the new physics on the defect, and we expect new
correlators of the form (\ref{One}), (\ref{CorrectedTwo}) to appear.
On the gravity side, this is a consequence of couplings of
closed-string modes to brane modes that are implicit in $S_{BI}$ and
$S_{WZ}$.  Furthermore, terms in the brane action involving open
string modes and open/closed string couplings make predictions for
purely three-dimensional correlators of the ${\cal O}_3$, as well as
mixed correlators involving both ${\cal O}_3$ and ${\cal O}_4$, which
we expect to match for example (\ref{Mixed}).

Let us compare the normalizations of the terms in (\ref{TotalAction})
to understand the relative coupling strength of the various kinds of
interaction.  The overall normalization of $S_{IIB}$ in Einstein frame
is \cite{Joe}
\begin{eqnarray}
\label{IIBAction}
S_{IIB} = \frac{1}{2 \kappa^2} \int d^{10}x \sqrt{-g}\left( R - \tf12
\, (\partial \Phi)^2 + \cdots \right) \,,
\end{eqnarray}
where $\kappa^2 = \tf12 (2 \pi)^7 g_s^2 {\alpha'}^4$ includes factors
of the string coupling extracted from the dilaton before passing to
Einstein frame.  In calculating correlation functions, it is useful to
Weyl-rescale the metric to extract the dimensionful parameter
\cite{BST},
\begin{eqnarray}
\label{WeylRescale}
g_{MN} \equiv L^2 \, \bar{g}_{MN} \,.
\end{eqnarray}
In terms of the rescaled metric, we have
\begin{eqnarray}
S_{IIB} \sim {L^8 \over g_s^2 {\alpha'}^4} \int d^{10}x
\sqrt{-\bar{g}} \left(R - \tf12 \, (\partial \Phi)^2 +
\cdots \right) \sim N^2 \int d^{10}x \sqrt{-\bar{g}} \left(R - \tf12
\,  (\partial \Phi)^2 + \cdots \right) ,
\label{NIIB}
\end{eqnarray}
where in the last line we used $L^4 = 4 \pi g_s N {\alpha'}^2$.  This
is a familiar result.  If we wish we may canonically normalize the
action by defining rescaled bulk fields $\Phi' \equiv \Phi N$.

The D5-brane action in Einstein frame is given by
\begin{eqnarray}
S_{BI} &=& - T_{D5} \int d^6\xi \,e^{\Phi / 2} \sqrt{- \det (g^{PB}_{ab} + e^{-\Phi/2} (B^{PB}_{ab} + 2 \pi \alpha ' F_{ab}))} \,, \\
S_{WZ} &=& - T_{D5} \int e^{2 \pi \alpha' F + B^{PB}} \wedge \sum_p C^{PB}_{(p)} \,, 
\end{eqnarray}
where $PB$ denotes the pullback of a ten-dimensional quantity; the
unusual powers of $e^{\Phi}$ result from transforming out of string
frame and do not affect the quadratic action. We use $a,b =
0,1,2,v,\theta,\varphi$ for the coordinates along the 5-brane, $i,j =
6, \chi,\varsigma,\psi$ for the normal directions, and $M,N$ to run
over all 10 indices.  Furthermore, we will use $\mu,\nu$ for $AdS_4$
indices alone and $\alpha, \beta$ for $S^2$ indices.
Weyl-rescaling the metric in  $S_{BI}$, we will find 
\begin{eqnarray}
\label{BIWeyl}
S_{BI} = - L^6 T_{D5} \int d^6 \xi {\sqrt{-\bar{g}}} (1 + {\rm
fluctuations}) = - \varrho N \lambda^{1/2} \int d^6 \xi
{\sqrt{-\bar{g}}} (1 + {\rm fluctuations}) \,,
\end{eqnarray}
where we used the expression $T_{D5} = 1 / (2 \pi)^5 g_s \,
{\alpha'}^3$ for the D5-brane tension, and $\varrho$ includes the
numerical factors.

\subsection{Correlators of dCFT from gravity}

Let us imagine a generic D5-brane field $\psi$ and some coupling of $m$ bulk generic fields $\phi$ to $n$ 5-brane fields:
\begin{eqnarray}
\label{Couplings}
S_{BI} &=& 
N \lambda^{1/2} \int d^6 \xi 
\left( (\partial \psi)^2 + \phi^m  \psi^n \right) \nonumber \\
&=& \int d^6 \xi \left( (\partial \psi')^2 + {1\over 
N^{m + n/2-1} \lambda^{n/4 -1/2}} \, {\phi'}^m {\psi'}^n \right).
\end{eqnarray}
where we defined a canonically normalized brane field $\psi' = N^{1/2}
\lambda^{1/4} \psi$.  The interaction terms resulting from $S_{WZ}$ 
scales identically in $N$ and $\lambda$. 
The canonically normalized fields $\phi'$ and $\psi'$ produce
two-point correlation functions of dual operators with no factors of
$N$ and $\lambda$.  With this normalization, the one-point function
of the bulk field $\phi'$ scales as $\lambda^{1/2}$
($m=1,n=0$) and the two-point function of the bulk field
and the defect field scales as $\lambda^{1/4} N^{-1/2}$
($m=1,n=1$).

Holography requires that the power of $N$ in the gravity
result for any correlator agree with that of planar graphs
in the field theory at fixed $\lambda$. On the other hand the
power of $\lambda$ from (\ref{Couplings}) at large $\lambda=g^2 N$ need not
agree with field theory results at weak coupling. It is quite
easy to see in the present case that the $N$-dependence always agrees
but the $\lambda$-dependence generically does not.

The agreement for $N$ can best be ascertained in the normalizations of
(\ref{NIIB}) and (\ref{BIWeyl}) in which we have the factor $N^2$ in
$S_{IIB}$ and $N$ in $S_{BI}$. All correlators $\langle {\cal O}_4\,
{\cal O}_4' \cdots \rangle$ which are non-vanishing if the defect is
removed are of leading order $N^2$, while contributions of $S_{BI}$
are of order $N$ in all correlators.  There is a simple normalization
in the dCFT which reproduces these results\footnote{For chiral
primaries one can take ${\cal O}_4 = N^{1- k/2} \, {\rm Tr} \, X^k$ in
terms of canonical $X$ fields of ${\cal N}=4$ SYM. Defect correlators
containing powers $(\bar{\Psi} \Psi)^j$ or $(\bar{q} q)^j$ of
canonical hypermultiplet fields carry the factor $N^{1-j}$.}. Planar
graphs with only adjoint fields are of order $N^2$, while in those with
defect fields there is a fundamental "quark" loop which matches the
$N$ in (\ref{BIWeyl}).

The power of $\lambda$ for multi-point correlators is generically a
negative fraction, and it is clear that perturbative field theory
gives non-negative integer powers in the weak coupling limit. This
situation is entirely consistent with the view that $AdS$/CFT
amplitudes sum all planar graphs at large fixed $\lambda$, but it also
indicates that the non-renormalization properties of correlation
functions in ${\cal N}=4$ SYM which were revealed through supergravity
\cite{LMRS} are absent for generic defect
correlators\footnote{O.~Aharony and A.~Karch independently calculated
the $\lambda$-dependence of $\langle {\cal O}_4 \rangle$ and
recognized it could not obey a non-renormalization theorem.  We thank
them for communicating their results.}.  Correlation functions with
$n=2$ and any $m$, however, are seen from (\ref{Couplings}) to be
independent of $\lambda$. This includes defect 2-point functions
$\langle {\cal O}_3 \, {\cal O}_3 \rangle$ and others which behave as
$\lambda^0$ at weak coupling.  Non-renormalization theorems could exist
for this class of correlators.

One can use (\ref{Couplings}) to compute correlation functions of
defect and ambient operators ${\cal O}_3$ and ${\cal O}_4$ for a
generic boundary dCFT.  The one-point function $\langle {\cal O}_4
\rangle$ is computed by taking the standard bulk-boundary propagator
in $AdS_5$, fixing a point on the boundary where ${\cal O}_4$ is
located, and integrating the propagator over the $AdS_4$ subspace.
Let us consider a scalar ${\cal O}_4$ of conformal weight $\Delta_4$.
The integral is convergent for $\Delta_4>3$, and one finds
\begin{eqnarray}\label{onepoint}
\langle {\cal O}_4(x,\vec{y}) \rangle&= & \lambda^{1/2} \int {dz \,
{d\vec{z}}^3\over z^4}{\Gamma(\Delta_4) \over \pi^2 \,
\Gamma(\Delta_4-2)} \left( {z \over z^2 + x^2 + (\vec{z}-\vec{y})^2 }
\right)^{\Delta_4} \nonumber \\ & = & \lambda^{1/2}{1 \over
x^{\Delta_4}} {\Gamma\left({\Delta_4-3\over 2}\right)
\Gamma\left({\Delta_4 \over 2}\right) \Gamma\left({3\over 2}\right)
\over \pi \, \Gamma(\Delta_4 -2)}.
\end{eqnarray}
By translational invariance along the defect, the one-point function
depends only on the transverse coordinate $x$.  The scaling
$x^{-\Delta_4}$ is what is expected from conformal invariance
(\ref{One}).  We will discuss the the singularity at $\Delta_4=3$
shortly.

The one-point function $\langle {\cal O}_4 (x, \vec{y}) \rangle$ is
closely related to the two-point function $\langle {\cal O}_4(x^\mu_1)
\, {\cal O}_4(x^\mu_2) \rangle$ in the conventional
$AdS_{d+1}$/CFT$_d$ correspondence.  It is known \cite{FMMR} that a
naive supergravity computation for the latter is incorrect and that a
careful cutoff procedure is required. One may thus be worried about a
similar sensitivity in the computation of $\langle {\cal O}_4
\rangle$.  However, there is reason to believe that this is not the
case here, and that (\ref{onepoint}) is in fact the correct answer.
One way to see this is to recall that for the two-point function, each
of the two contributing terms from the action was separately
divergent, and so a more careful treatment of the Dirichlet problem
was required to extract the proper finite result \cite{FMMR}.  Here
there is no such divergence in the single term contributing to the
one-point function.

Alternately, a worldsheet way to understand the subtlety in the
computation of the two-point function follows from trying to perform
the calculation in string theory, which is well-defined for $d=2$
\cite{MO}.  There one considers a two-point function of the
corresponding vertex operators on a sphere and divides the result by
the volume of the worldsheet conformal symmetry which fixes the two
insertion points. The volume of this residual conformal symmetry is
infinite, and it is canceled by another infinity in the numerator
from the worldsheet two-point function.  Thus again the computation of
the target space correlator involves cancellation of two divergent
factors, which may leave out a finite $\Delta$-dependent coefficient;
in fact the proper treatment of this computation has been shown to
give the correct factor for $d=2$ \cite{MO}.  However, there is no
corresponding subtlety in the computation of the one point function
$\langle {\cal O}_4 \rangle$, since the volume of the residual
conformal symmetry of a disk with one interior point fixed is
finite. Hence we expect (\ref{onepoint}) to be unambiguous and correct.

For the two-point function $\langle {\cal O}_4(x,\vec{y}) {\cal
O}_3(\vec{0}) \rangle$, the integral to be done is the product of
bulk-boundary propagators $K_{\Delta_4} K_{\Delta_3}$, with the first
as above and the second propagating from the point $z_\mu=(z,0,\vec{z})$
on $AdS_4$ to the point $\vec{0}$ on its boundary. We write
\begin{eqnarray}
\label{twopoint}
\langle {\cal O}_4(x,\vec{y}) \, {\cal O}_3(\vec{0})\rangle
= J(x,\vec{y}; \Delta_4, \Delta_3)\, {\lambda^{1/4} \over
N^{1/2}}{\Gamma(\Delta_4) \over \pi^2\Gamma(\Delta_4-2)}
{\Gamma(\Delta_3) \over\pi^{3/2}\Gamma(\Delta_3-{3\over2})} \,,
\end{eqnarray}
with the integral
\begin{eqnarray}
J(x,\vec{y}; \Delta_4, \Delta_3) = \int{dz \, {d\vec{z}}^3\over z^4}\left({z \over z^2 + x^2 +
(\vec{z}-\vec{y})^2}\right)^{\Delta_4} \left({z \over z^2 + \vec{z}^2}\right)^{\Delta_3} \,.
\end{eqnarray}
As explained in \cite{FMMR}, it is convenient to use the inversion
\begin{eqnarray}
(z,0,\vec{z}) \equiv {1 \over z'^2+ \vec{z}'^2} (z',0,\vec{z}') \,,
\quad \quad (x,\vec{y}) \equiv {1 \over x'^2+ \vec{y}'^2}
(x',\vec{y}')\,,
\end{eqnarray}
to do the integral, which leads to
\begin{eqnarray}
J ={1 \over (x^2 + \vec{y}^2)^{\Delta_4}} \int dz'd\vec{z}'^3
 (z')^{\Delta_3-4}\left({z '\over z'^2 + x'^2 + (\vec{z'}-\vec{y'})^2
 }\right)^{\Delta_4} \,.
\end{eqnarray}
After scaling $\vec{z}' = \vec{y}' +\sqrt{x'^2+z'^2}\vec{w}$ and $z'=x' u$, 
one finds
\begin{eqnarray}
\label{Jint}
J&=&{1 \over x^{\Delta_4-\Delta_3} (x^2 + \vec{y}^2)^{\Delta_3}} \int
{du \, u^{\Delta_4+\Delta_3-4} \over (1 +u^2)^{\Delta_4 -{2\over2}}}
\int {d\vec{w}^3 \over (1+ \vec{w})^{\Delta_4}} \\ &=& {1 \over
x^{\Delta_4-\Delta_3} (x^2 + \vec{y}^2)^{\Delta_3}} { \pi^2
\Gamma({\Delta_4+\Delta_3-3 \over 2}) \Gamma({\Delta_4-\Delta_3 \over
2}) \over {2 \Gamma(\Delta_4)}} \,.
\end{eqnarray}
The conformal invariant form (\ref{Mixed}) thus arises from gravity. The
integral converges if the conditions $\Delta_4 \ge \Delta_3$ and
$\Delta_4+\Delta_3 \ge 3$ are satisfied. The singularity at
$\Delta_4+\Delta_3 =3$ is due to a divergence as the inverted radial
coordinate $z' \rightarrow 0$ and is similar to the singularity of
$\langle {\cal O}_4 \rangle$ at $\Delta_4=3$.  The singularity at $
\Delta_4 =\Delta_3$ arises as $z' \rightarrow \infty$.

The poles due to the $\Gamma$-functions in the numerators of
(\ref{onepoint}) and (\ref{Jint}) were calculated using the generic
form (\ref{Couplings}) of $S_{BI}$. We can show that they cancel in
the particular $D3/D5$ theory we are studying because the actual
couplings vanish due to $SU(2)_H \times SU(2)_V$ symmetry. For
(\ref{onepoint}) the issue arises only $\Delta_4=3$, but the primary
operator $ {\cal O}_4= {\rm Tr}\ X^3$ belongs to the $(0,3,0)$
irreducible representation of $SO(6)$ which contains no singlets under
the residual R-symmetry.

To discuss the poles in $\langle{\cal O}_4 \, {\cal O}_3 \rangle$, we
must anticipate one key result of the Kaluza-Klein analysis in the
next subsection, namely that the primary operators on the defect carry
$SU(2)_H\times SU(2)_V$ quantum numbers $(\ell \ge 1,0)$ and have
scale dimension $\Delta_3 =\ell$. Thus the pole at $\Delta_4+\Delta_3
= 3$ in (\ref{Jint}) can appear only for ${\cal O}_4={\rm Tr}\, X^2$
and the lowest ${\cal O}_3$, a case which violates
R-symmetry. Consider next poles at $\Delta_3-\Delta_4 = 2n \ge 0$. We
need the fact that the primaries ${\cal O}_4={\rm Tr}\, X^k$ contain
only components in the representations $(k,0),(k-2,0), \cdots $ of
$SU(2)_H\times SU(2)_V$.  Isospin conservation in $\langle{\cal O}_4
\, {\cal O}_3 \rangle$ thus requires $\ell=k-2m$ or $\Delta_3-\Delta_4
= \ell -k =-2m$; thus only the case with pole $\Delta_3-\Delta_4=0$ is
allowed by R-symmetry. However, the set of poles we are discussing are
close analogues of those in the 3-point function on ${\cal N}=4$ SYM
$\langle {\rm Tr}\, X^k\, {\rm Tr}\, X^l \, {\rm Tr} \,X^m\rangle$
studied in \cite{FMMR}. In the 3-point case a large set of singular
cases is forbidden by $SO(6)$ symmetry, and there is one remaining
extremal case with $k=l+m$. For this case the actual bulk couplings
$g_{klm}$ from Type IIB supergravity have a zero which cancels the
pole leaving a finite result \cite{LMRS}. The remaining singular case
for $\langle{\cal O}_4 \, {\cal O}_3 \rangle$ is extremal in exactly
the same sense, and we conjecture that the specific couplings that
occur in the $D5$-brane action will cancel the pole.

\subsection{D5-brane open-string modes}

We now turn to a more detailed study of the D5-brane action for the
Karch-Randall system.  We will enumerate all terms up to quadratic
order in both open and closed string bosonic fluctuations.
Considering first the quadratic action for the open string modes
alone, we perform a Kaluza-Klein reduction on the $S^2$, producing
kinetic terms for towers of $AdS_4$ modes.  We solve for the masses of
these fluctuations, and determine the conformal dimensions of the dual
operators ${\cal O}_3$.  As we will see, two kinds of excitation are
elementary to handle, while the remaining two types are mixed and
their mass matrix must be diagonalized.  Although there are three
negative-mass modes in the full system, the Breitenlohner-Freedman
stability bound is satisfied. All masses and conformal dimensions are
nontrivially found to be integers, a sign of supersymmetry.  These
fluctuations fit into short multiplets of $OSp(4|4)$, and we will
establish the dictionary relating them to gauge-invariant defect
operators in the dual dSCFT in section~\ref{OperatorSec}.

There still remain interactions on the brane involving closed string
modes.  As explained in the last subsection, these give rise to
various correlation functions.  We list the couplings up to
quadratic order in section \ref{ClosedSec}, but do not perform the KK
reductions for most cases.

The bosonic open string modes living on the D5-brane are the $U(1)$
gauge field $B_a$ and the embedding coordinates $Z^M$.\footnote{We
reserve the symbols $A$ and $X$ for the D3-brane fields that will
appear in the field theory sections.  $B_a$ should not be confused
with the NSNS 2-form $B_{ab}$.}  As usual we pick a static gauge to
fix the worldvolume diffeomorphisms, $\xi^a = Z^a$, leaving us with
the dynamical fluctuations $Z^i$. Expanding out the determinant in
$S_{BI}$ to quadratic order, we find
\begin{eqnarray}
\label{BIExpand}
S_{BI} = - T_{D5} \int d^6 \xi \, e^{\Phi/2} \sqrt{-\det{g}} \left(1 + {1 \over 2} \partial^a Z^i \partial_a Z^j g_{ij} + {1 \over 4} {\cal F}_{ab} {\cal F}^{ab} + \partial^a Z^i h_{ia} \right) \,,
\end{eqnarray}
where ${\cal F}_{ab} \equiv B_{ab} + 2 \pi \alpha' F_{ab}$.  There is
still a lot of physics hidden in $\sqrt{- \det{g}}$, which is the
determinant of the metric over the $AdS_4 \times S^2$ directions.  The
background metric is implicitly a function both of the worldvolume
coordinates $\xi^a$ (thanks to the static gauge condition) and the
embedding fields $Z^i$:
\begin{eqnarray}
\sqrt{- \det g}= L^6 v^2 \sin \theta \cos^2 Z^\psi =  L^6 \sqrt{-\bar{g}_4} \, \sqrt{\bar{g}_{2}} \cos^2 Z^\psi\,,
\end{eqnarray}
where $\bar{g}_4$ and $\bar{g}_2$ are the determinants of the
Weyl-rescaled metric (\ref{WeylRescale}) on $AdS_4$ and $S^2$,
respectively.  The cosine will provide mass terms for $Z^\psi$.
Furthermore, $\bar{g}_4$ and $\bar{g}_2$ contain graviton fluctuations,
which must be expanded out when we consider closed string modes.

For now, we concentrate on the open string modes in (\ref{BIExpand})
and postpone discussing the closed string fluctuations, including
those in mixed terms such as $\partial^a Z^i h_{ia}$.  For the various
$g_{ij}$, we find
\begin{eqnarray}
g_{xx} = L^2 v^2 \,, \quad g_{\psi \psi} = L^2 \,, \quad
g_{\chi \chi} = L^2 \sin^2 Z^\psi \,, \quad
g_{\varsigma \varsigma} = L^2 \sin^2 Z^\psi \sin^2 Z^\chi \,.
\end{eqnarray}
Notice that $g_{\chi \chi}$ and $g_{\varsigma \varsigma}$ are higher
order in the fluctuations, and hence the kinetic terms for $Z^\chi$
and $Z^\varsigma$ vanish to quadratic order.  This is a consequence of
our choice of coordinate system, as the $\chi$ and $\varsigma$
coordinates become degenerate at $\psi = 0$, the location of the
D5-brane.  All infinitesimal fluctuations of the D5-brane on the
$S^5$ are $\psi$ fluctuations, and they form a triplet of $SU(2)_V$.
Thus $S_{BI}$ to quadratic order in open string fluctuations reads
\begin{eqnarray}
\label{QuadBI}
S_{BI} = - (T_{D5} L^6) \int d^4x \sqrt{\bar{g}_4} \!\!&\!\! d\Omega
\!\!&\!\! (1 + \tf12 v^2 \partial^a Z^x \partial_a Z^x \\ &+& \tf12
\partial^a Z^\psi \partial_a Z^\psi - (Z^\psi)^2 + \left(\tf{2 \pi
{\alpha'}}{L^2}\right)^2 \tf14 F_{ab} F^{ab} ) \,, \nonumber
\end{eqnarray}
where we are now raising indices with $\bar{g}^{ab}$, and $d \Omega
\equiv \sqrt{\bar{g}_2} d\theta d\varphi$.  Notice that the gauge
field kinetic term is down by an additional factor ${\alpha'}^2 / L^4
\sim 1/\lambda$.

Let us now turn to $S_{WZ}$.  We find
\begin{eqnarray}
\label{WZExpand}
S_{WZ} = - T_{D5} \int \left( C_6^{PB} + C_4^{PB} \wedge {\cal F} + \cdots \right) \,.
\end{eqnarray}
Of the Ramond-Ramond fields, only $C_4$ is nonzero in the background.
The relevant term\footnote{There is also a term polarized in the
angular directions, required for the self-duality of $F_5$; it does
not play a role in the quadratic Lagrangian.} is
\begin{eqnarray}
\label{BackgdC}
C_{x012} = v^4 L^4 \,.
\end{eqnarray}
The 5-brane does not span the coordinate $x$.  However,
(\ref{BackgdC}) contributes to the pullback
\begin{eqnarray}
C^{PB}_{abcd} = \partial_a Z^i C_{ibcd} + ( {\rm perms\ in\ } abcd) + {\cal O}(Z^2) \,.
\end{eqnarray}
We find the contribution to the part of $S_{WZ}$ quadratic in
five-brane fields,
\begin{eqnarray}
\label{QuadWZ}
S_{WZ} &=&  - \frac12 L^4 T_{D5} (2 \pi \alpha') \int d^6 x \, v^4 \,
\tilde\epsilon^{\alpha \beta} (2 \partial_\alpha Z^x F_{v \beta} -
F_{\alpha \beta} \partial_v Z^x ) \,, \\ &=& - \frac12 L^4 T_{D5}(2
\pi \alpha') \int d^6 x \, v^4 \, \tilde\epsilon^{\alpha \beta} (2
\partial_\alpha Z^x \partial_v B_\beta - F_{\alpha \beta} \partial_v
Z^x ) \,, \nonumber
\end{eqnarray}
where $\tilde{\epsilon}^{\alpha \beta}$ is the flat-space epsilon
tensor with $\tilde{\epsilon}^{\theta \phi} = 1$, and we used integration by
parts and antisymmetry to eliminate the
$\tilde{\epsilon}^{\alpha\beta} \partial_\alpha Z^x \partial_\beta
B_v$ term in the second line of (\ref{QuadWZ}). Combining
(\ref{QuadBI}) and (\ref{QuadWZ}), we have the complete set of
quadratic terms in the open-string fields.  We see that the gauge
field is coupled to the scalar $Z^x$, while the scalar $Z^\psi$ is
free.  We examine each of these systems in turn, expanding in
spherical harmonics on the $S^2$ and computing $AdS_4$ masses and dual
conformal dimensions.

\medskip \noindent \underline{\bf Angular fluctuations} \quad \quad
The D5-brane may wiggle away from its background location $\psi = 0$
on the 5-sphere, and this is described by $Z^\psi$.  The fluctuation
equation follows from (\ref{QuadBI}) and is simply
\begin{eqnarray}
\label{PsiEqn}
\left(\, \square + \, 2 \right) \, Z^\psi = 0 \,. 
\end{eqnarray}
We expand in the usual $S^2$ spherical harmonics,
\begin{eqnarray}
\label{PsiY}
Z^\psi(\vec{y}, v, \Omega) = \sum_{l,m} \psi^l_m(\vec{y}, v) Y^l_m(\Omega) \,.
\end{eqnarray}
The six-dimensional Laplacian splits as $\square = \square_{AdS_4} +
\square_{S^2}$, and as every second-grader knows from studies of
angular momentum, the spherical harmonics $Y^l_m(\theta,\phi)$ are
eigenvectors of $\square_{S^2}$ with eigenvalues
\begin{eqnarray}
\square_{S^2} Y^l_m(\theta, \varphi) = - l (l+1) \, Y^l_m(\theta, \varphi) \,.
\end{eqnarray}
Upon reduction, (\ref{PsiEqn}) becomes
\begin{eqnarray}
\label{PsiMasses}
\left( \square_{AdS_4} - m^2(l) \right) \psi^l_m(x) = 0 \,, \quad \quad m^2(l) = -2 + l (l+1) \,. 
\end{eqnarray}
Thus the zero mode is tachyonic.  However, tachyonic modes do not
generate an instability in $AdS_{d+1}$ space as long as the masses do
not violate the Breitenlohner-Freedman bound \cite{BF}, which in the
metric $\bar{g}$ where the $AdS$ scale is unity takes the form $m^2
\geq - d^2/4$.  For $d=3$ we have $m^2 \geq -9/4$, which is satisfied
by all the modes (\ref{PsiMasses}).  Hence there is no instability in
this sector, as expected due to supersymmetry.  Karch and Randall
\cite{KR1} already considered the zero mode and found it to be stable.

Using the standard $AdS_{d+1}$/CFT$_{d}$ formula $\Delta_{\pm} = (d \pm \sqrt{ d^2 + 4 m^2}) /2$ with $d=3$, we find for the dual conformal dimensions,
\begin{eqnarray}
\Delta_+ = 2 + l \,, \quad \quad \Delta_- = 1 - l \,.
\label{AngularDelta}
\end{eqnarray} 
$\Delta_-$ is only possible for the constant mode $l=0$.

\medskip \noindent \underline{\bf $AdS_4$ gauge field fluctuations}
\quad \quad We find it convenient to define $b_a \equiv (2 \pi \alpha'
/ L^2) B_a$, $f_{ab} \equiv (2 \pi \alpha' / L^2) F_{ab}$; these
fluctuations then have the same normalization as the $Z^i$. The action
is then
\begin{eqnarray}
\label{GaugeAction}
S_{gauge} &=& - {1 \over 4} T_{D5} L^6 \int d^4x \sqrt{\bar{g}_4} \, d \Omega \, f_{ab} f^{ab}
  \\ &=& - {1 \over 4} T_{D5} L^6 \int d^4x \sqrt{\bar{g}_4} \, d \Omega \, ( f_{\mu\nu} f^{\mu\nu} + 2 f_{\mu \alpha} f^{\mu \alpha}  + f_{\alpha \beta} f^{\alpha \beta} ) \,.
\nonumber
\end{eqnarray}
We impose the gauge choice $D^\alpha b_\alpha = 0$, which decouples
$b_\mu$ from $b_\alpha$.  We then find for $S_{gauge}$
\begin{eqnarray}
S_{gauge} &=& S_{b_\mu} + S_{b_\alpha} \,, \\
S_{b_\mu} &=&  - {1 \over 4} T_{D5} L^6 \int d^4x \sqrt{\bar{g}_4} \, d \Omega \, ( f_{\mu\nu} f^{\mu\nu} + 2 D_\alpha b_\mu D^\alpha b^\mu ) \,, \\
S_{b_\alpha} &=&  - {1 \over 4} T_{D5} L^6 \int d^4x \sqrt{\bar{g}_4} \, d \Omega \, ( 2 D_\mu b_\alpha D^\mu b^\alpha + f_{\alpha \beta} f^{\alpha \beta} ) \,.
\label{Salpha}
\end{eqnarray}
Furthermore, we see that the coupling (\ref{QuadWZ}) involves only
$b_\alpha$.  Therefore $S_{b_\mu}$ gives the complete quadratic action
for $b_\mu$.  The fluctuation equation is
\begin{eqnarray}
\label{fEqn}
D^\mu f_{\mu \nu}  + \square_{S^2} b_\nu = 0\,.
\end{eqnarray}
The $b_\mu$ are scalars on the $S^2$ and hence can be expanded in ordinary spherical harmonics as with (\ref{PsiY}),
\begin{eqnarray}
\label{bY}
b_\mu(\vec{y}, v, \Omega) = \sum_{l,m} b^l_{\mu m}(\vec{y}, v) Y^l_m(\Omega) \,,
\end{eqnarray}
under which (\ref{fEqn}) reduces to a Maxwell equation for the zero-mode
and standard Proca equations for the excited tower, with masses
\begin{eqnarray}
\label{VectorMasses}
m^2 = l (l+1) \,.
\end{eqnarray}
We translate (\ref{VectorMasses}) into conformal dimensions for dual
operators using the standard vector relation $\Delta = (d + \sqrt{
(d-2)^2 + 4 m^2})/2$, and obtain
\begin{eqnarray}
\label{GaugeDelta}
\Delta = 2 + l \,.
\end{eqnarray}  

\medskip \noindent \underline{\bf Coupled sector} \quad \quad We
finally consider the coupled sector of $b_\alpha$ and $Z^x$ from
(\ref{QuadBI}), (\ref{QuadWZ}), and (\ref{Salpha}).  In this instance
we find it more convenient to perform the $S^2$ reduction on the level
of the action, before extracting equations of motion for each mode.

For $Z^x$ we expand as usual
\begin{eqnarray}
\label{ZY}
Z^x(\vec{y}, v, \Omega) = \sum_{l,m} z^l_m(\vec{y}, v) Y^l_m(\Omega) \,.
\end{eqnarray}
For $b_\alpha$, the gauge condition $D^\alpha b_\alpha =0 $ tells us
that $b$ is co-closed as a 1-form on $S^2$; by the Hodge decomposition
theorem $b$ is a sum of co-exact and harmonic pieces.  Since there are
no harmonic 1-forms on $S^2$, we may write $b$ as a co-exact form, 
\begin{eqnarray}
\label{alphaY}
b_\alpha(\vec{y}, v, \Omega) = \sum_{l,m} b^l_m(\vec{y}, v)
\,\epsilon_{\alpha \beta} D^\beta Y^l_m(\Omega) \,,
\end{eqnarray}
where $\epsilon_{\alpha \beta}$ is the curved-space epsilon-tensor on
$S^2$.  In what follows, we will drop the ``magnetic quantum number''
$m$ on $z^l_m$, $b^l_m$ and $Y^l_m$ for clarity; it is implicitly
present and summed over when $l$ is summed over.

We find in (\ref{Salpha}),
\begin{eqnarray}
\int d \Omega \,  2 D_\mu b_\alpha D^\mu b^\alpha &=& 2 \int d \Omega \, \sum_{l l'} \left( D_\mu b^l D^\mu b^{l'} \right) \left( D_\beta Y^l D^\beta Y^{l'} \right) \,, \\
&=& 2 k(l) \sum_l {l (l+1) \over L^2} D_\mu b^l D^\mu b^l \,,  
\nonumber
\end{eqnarray}
where we integrated by parts and used (\ref{alphaY}), and $k(l)$
is the normalization in $\int d\Omega Y^l Y^{l'} = k(l) \delta^{l l'}$, which will drop out at the end of the day, as well as
\begin{eqnarray}
\int d\Omega \, f_{\alpha \beta} f^{\alpha \beta} &=&
2 \int d\Omega  \, \sum_{l l'} \, b^l \, b^{l'} \left[ (D_\alpha \epsilon_{\beta \gamma} - D_\beta \epsilon_{\alpha \gamma} )  D^\gamma Y^l \right]
D^\alpha \epsilon^{\beta \delta} D_\delta Y^{l'} \,, \\
&=& 2 k(l) \sum_l (l (l+1))^2 \, b^l \, b^l \,, 
\nonumber
\end{eqnarray}
where we have commuted covariant derivatives through each other as
needed and used $\bar{R}_{\alpha \beta} = \bar{g}_{\alpha \beta}$.
Thus the total action (\ref{Salpha}) for the $b^l$ modes is
\begin{eqnarray}
S_{b_\alpha} = - {1 \over 4} T_{D5} L^6 \, k(l) \int d^4x \sqrt{\bar{g}_4} \,
2 \sum_l l (l+1) \left( \partial_\mu b^l \partial^\mu b^l + l (l+1) \, b^l \, b^l \right) \,.
\label{ReducedA}
\end{eqnarray}
The quadratic terms for $Z^x$ in (\ref{QuadBI}) are considerably simpler; we find
\begin{eqnarray}
\nonumber
S_x &=& - {1 \over 2} T_{D5} L^6 \int d^4x \sqrt{\bar{g}_4} \, d \Omega \,v^2 \sum_{l l'} \left( D_\mu z^l D^\mu z^{l'} Y^l Y^{l'} + z^l z^{l'} D^\alpha Y^l D_\alpha Y^{l'} \right) \,, \\  
&=& - {1 \over 4} T_{D5} L^6 \, k(l) \int d^4x \sqrt{\bar{g}_4} \,  (2 v^2) \sum_{l} \left( \partial_\mu z^l \partial^\mu z^{l} + l (l+1) \, z^l z^{l}\right) \,.
\label{ReducedX}
\end{eqnarray}
Finally, there is the mixing term from $S_{WZ}$ (\ref{QuadWZ}).
Writing $\tilde{\epsilon}^{\alpha \beta} f_{\alpha \beta} = 2
\tilde{\epsilon}^{\alpha \beta} \partial_\alpha a_\beta$, we integrate
both the $\partial_\alpha$ and the $\partial_v$ derivatives in the
second term in (\ref{QuadWZ}) by parts, which cancels the first term but
leaves a piece coming from $(\partial_v v^4)$.  Using $\tilde{\epsilon}^{\alpha \beta} = \sqrt{\bar{g}_2} \epsilon^{\alpha \beta}$ and a factor of $v^2$
to form $\sqrt{\bar{g}_4}$, we obtain
\begin{eqnarray}
S_{WZ} = - 4 T_{D5} L^6 \int d^4x \sqrt{\bar{g}_4} \, d \Omega \, v Z^x  \epsilon^{\alpha \beta} D_\alpha b_\beta \,.
\end{eqnarray}
Expanding both $Z^x$ and $b_\alpha$ in spherical harmonics, we find
\begin{eqnarray}
\nonumber
S_{WZ} &=&  4 T_{D5} L^6 \int d^4x \sqrt{\bar{g}_4} d \Omega \, v \sum_{l l'} z^l Y^l b^{l'} \square_{S^2} Y^{l'} \,, \\
&=& - {1 \over 4} T_{D5} L^6 k(l) \int d^4x \sqrt{\bar{g}_4} \sum_l 16 \, l (l+1)  \, v z^l \, b^l \,.
\label{ReducedWZ}
\end{eqnarray}
We are now in a position to derive the fluctuation equations for each
mode using the total action (\ref{ReducedA}), (\ref{ReducedX}),
(\ref{ReducedWZ}).  For the $b^l$ modes, we find
\begin{eqnarray}
\label{AEqn}
\square_{AdS_4} b^l = l (l+1) \,b^l + 4 v z^l \,,
\end{eqnarray}
while for the $z^l$, we have
\begin{eqnarray}
\label{XEqn}
 {1 \over \sqrt{\bar{g}_4}} \partial_\mu \sqrt{\bar{g}_4} \, v^2 g^{\mu\nu} \partial_\nu z^l = v^2 l(l+1) z^l + 4 l (l+1)\, v \, b^l \,,
\end{eqnarray}
The factors of $v^2$ can be dealt with by rescaling $z^l$ by a
function of $v$ that is chosen to eliminate any terms with a single
derivative of $z^l$ on the left-hand side of (\ref{XEqn}).  The
correct factor to extract turns out to be
\begin{eqnarray}
y^l \equiv v z^l \,.
\end{eqnarray}
Dividing by an overall factor of $v$, equation (\ref{XEqn}) then reduces to
\begin{eqnarray}
\label{XEqn2}
(\square_{AdS_4} - 4 ) \, y^l = l(l+1) \, y^l + 4 l(l+1) \, b^l \,.
\end{eqnarray}
Additionally, the equation for $b^l$ (\ref{AEqn}) loses its explicit
factors of $v$ when expressed in terms of $y^l$:
\begin{eqnarray}
\label{AEqn2}
\square_{AdS_4} b^l = l(l+1) \, b^l + 4 y^l \,.
\end{eqnarray}
Solving the system is now trivial.  The equations (\ref{XEqn2}),
(\ref{AEqn2}) can be expressed in terms of the mass matrix
\begin{eqnarray}
\label{MassMatrix}
  \square_{AdS_4} \pmatrix{ y^l \cr b^l} = 
  \pmatrix{ l(l+1) + 4 & 4l(l+1) \cr
   4  & l(l+1)}
  \pmatrix{ y^l \cr b^l } \,.
\end{eqnarray}
The mass matrix is diagonalized to find the mass eigenvalues
\begin{eqnarray}
m^2  &=& l(l+1) + 2 \pm 2 \sqrt{4 l(l+1) + 1} \,, \\
&=& l^2 + l + 2 \pm 2(l+1) \,.
\nonumber
\end{eqnarray}
The masses turn out integer, which is not a property of generic
Freund-Rubin-type KK reductions and is usually an indication of
supersymmetry \cite{DFGHM}.  Each of the two branches
\begin{eqnarray}
\label{Branches}
m^{2(+)} = l^2 +5l +4 \,, \quad \quad m^{2(-)} = l^2 -3l \,,
\end{eqnarray}
has associated dual operators, whose conformal dimensions we compute.
For $m^{2(+)}$, we have
\begin{eqnarray}
\Delta^{(+)}_\pm \;\; = \;\; {3 \over 2} \pm {1 \over 2} \sqrt{9 + 4 (l^2 + 5l +4)} \;\; = \;\; {3 \over 2} \pm {1 \over 2} (2l +5) \,.
\label{UpperBranch}
\end{eqnarray}
Only the $+$-branch is possible for unitarity; this gives
\begin{eqnarray}
\Delta^{(+)}_+ = l + 4 \,.
\end{eqnarray}
Meanwhile, for $m^{2(-)}$, we find
\begin{eqnarray}
\label{LowerBranch}
\Delta^{(-)}_\pm \;\; = \;\; {3 \over 2} \pm {1 \over 2} \sqrt{9 + 4 (l^2 -3l)} \;\; = \;\; {3 \over 2} \pm {1 \over 2} |2l -3| \,.
\end{eqnarray}
For $l=1,2$, both choices are possible, while only $\Delta^{(-)}_+$ is
possible for $l > 2$. Again nontrivially, we find integer quantities.

A few words are necessary for the special case $l=0$.  This
corresponds to a constant spherical harmonic $Y^{l=0}$.  It is easy to
see from (\ref{alphaY}) that $b_\alpha$ vanishes for such a mode, and
hence $b^{l=0} = 0$ uniformly.  (The expansion of the vector field
$b_\alpha$ on $S^2$ does not contain a scalar part.)  As a result the
$y^{l=0}$ mode is uncoupled, and from (\ref{XEqn2}) we see that it has
the (positive) mass $m^2 = 4 $.  This is merely the value of
$m^{2(+)}$ for $l=0$ (\ref{Branches}).  Hence, as is common in such
Kaluza-Klein problems, the lower branch truncates at some $l >0$, in
this case $l = 1$, while the upper branch can take any value $l \geq
0$.  The $l=1$, $l=2$ states on the lower branch both have the negative
mass $m^2 = - 2$, which satisfies the Breitenlohner-Freedman bound.

We have now determined the complete spectrum of bosonic open-string
fluctuations on the D5-brane.  These modes are expected to be the
bosonic elements of a series of short representations of the
superalgebra $OSp(4|4)$ whose even subalgebra is $SO(3,2) \times
SU(2)_H \times SU(2)_V$.  The structure of such representations is
known \cite{Kac}, but it is simpler to compare with the short
representations of maximum spin 1 of the $OSp(3|4)$ subalgebra whose
decomposition with respect to $SO(3,2)\times SO(3)$ was explicitly
given in (50) of \cite{Nicolai}.  The supercharges of $OSp(3|4)$ are
in the $J=1$ of $SO(3)$, so we identify $SO(3)$ as $SU(2)_D$, the
diagonal subalgebra of $SU(2)_H \times SU(2)_V$.  This means that the
$\psi$ modes appear with $J=l+1, l,l-1$. Having noted this, one finds
complete agreement between the Kaluza-Klein modes
(\ref{AngularDelta}), (\ref{GaugeDelta}), (\ref{UpperBranch}),
(\ref{LowerBranch}) and the short representations of \cite{Nicolai}.
Agreement for the bosonic modes is non-trivial since a given
$OSp(3|4)$ representation includes 5 scalars and a vector with
specific relations between $\Delta$ and $J$.  The KK spectrum is
summarized in Table 1 of section~\ref{OperatorSec}, where we will
match the D5-brane modes to gauge-invariant composite operators
confined to the defect of the dual field theory.

\subsection{D5-brane closed-string modes}
\label{ClosedSec}

Here we briefly list the remaining quadratic terms in the Born-Infeld
and Wess-Zumino actions, involving closed as well as open string
modes.  These generate $\langle {\cal O}_4 \rangle$, $\langle {\cal
O}_4 {\cal O}_3 \rangle$ and corrections to $\langle {\cal O}_4 {\cal
O}'_4 \rangle$, respectively.  We perform the KK reduction for the
example of the dilaton one-point coupling.

The Born-Infeld action (\ref{BIExpand}) contains terms involving the
graviton $h$ and dilaton $\Phi$.  Expanding the dilaton exponential and using
\begin{eqnarray}
\sqrt{g} = \sqrt{g^0} \left( 1 + \tf12 h^a_a + \tf18 (h^a_a)^2 - \tf14 h_{ab}h^{ab} + {\cal O}(h^3) \right) \,,
\end{eqnarray}
we find the closed-string one-point couplings,
\begin{eqnarray}
\label{OnePoint}
S_{BI}^{(1)} =- T_{D5} L^6 \int d^4x \sqrt{\bar{g}_4} \, d \Omega \left( \tf12 \Phi + \tf12 h^a_a \right) \,,
\end{eqnarray}
the closed-string two point-couplings 
\begin{eqnarray}
\label{TwoPoint}
S_{BI}^{(2)} =- T_{D5} L^6 \int d^4x \sqrt{\bar{g}_4} \, d \Omega \left(
\tf18 \Phi^2 + \tf18 (h^a_a)^2 - \tf14 h_{ab}h^{ab}
 + \tf14 \Phi h^a_a + \tf14 B_{ab} B^{ab} \right) \,,
\end{eqnarray}
and the mixed open/closed couplings
\begin{eqnarray}
\label{MixedPoint}
S_{BI}^{(1,1)} =- T_{D5} L^6 \int d^4x \sqrt{\bar{g}_4} \, d \Omega \left(
\partial^a Z^i h_{ia} + \tf14 B_{ab} f^{ab} \right) \,.
\end{eqnarray}
The Wess-Zumino action (\ref{WZExpand}) couples the closed-string
fluctuations $C_6$ and $C_4$ to the brane. The one-point coupling is
\begin{eqnarray}
\label{OnePointWZ}
S_{WZ}^{(1)} = 
-T_{D5} \int C_6 =
-T_{D5} L^6 \int d^4x \sqrt{\bar{g}_4} \, d \Omega  \, \left( \tf{1}{6!} \, \epsilon^{abcdef} (C_6)_{abcdef} \right)\,,
\end{eqnarray}
the closed string two-point coupling is
\begin{eqnarray}
\label{TwoPointWZ}
S_{WZ}^{(2)} = - T_{D5} \int B \wedge C_4 =
-T_{D5} L^6 \int d^4x \sqrt{\bar{g}_4} \, d \Omega  \, \left( \tf{1}{2 \cdot 4!} \, \epsilon^{abcdef} B_{ab} \, (C_4)_{cdef} \right)\,,
\end{eqnarray}
and the mixed two-point couplings are
\begin{eqnarray}
S_{WZ}^{(1,1)} = &-& T_{D5} L^6 \int d^4x \sqrt{\bar{g}_4} \, d \Omega \, [ \epsilon^{abcdef} \, \left(
\tf{1}{5!} \, (\partial_a Z^i) (C_6)_{ibcdef} + \tf{1}{2 \cdot 4!} \, f_{ab} \, (C_4)_{cdef} \right) \nonumber  \\ 
&& - \tf12 v^2 \epsilon^{\alpha \beta} 
\left(B_{\alpha \beta} \partial_v Z^x - 2 \partial_\alpha Z^x B_{v \beta} \right) ] \,. \label{MixedPointWZ}
\end{eqnarray}
Again, $\epsilon$ denotes a curved-space antisymmetric tensor.
Note the terms in the last line of (\ref{MixedPointWZ}) involved the
background value of $C_4$ and are analogous to the purely open-string
terms (\ref{QuadWZ}).

One point to notice is that the brane interactions do not couple bulk
eigenmodes directly.  Thus we find the one-point coupling
$h^\alpha_\alpha$ with $\alpha = \theta, \varphi$ in (\ref{OnePoint}),
although the field theory operators ${\rm Tr}\, X^k$ are dual to linear
combinations of $h^\alpha_\alpha$ with $\alpha$ now running over all
$S^5$ indices, and the four-form $C_4$ polarized along the $S^5$.

Naturally, all the bulk modes appearing in
(\ref{OnePoint})-(\ref{MixedPointWZ}) are restricted to the brane.
This implies certain restrictions on the $SO(6)$ quantum numbers of
the modes resulting from the $S^5$ reduction.  Consider the dilaton,
which is the simplest case since it is a 10D scalar.  As usual it is
expanded in spherical harmonics on $S^5$,
\begin{eqnarray}
\Phi(\vec{y}, x, v, \Omega_5) = \sum_I  
\Phi^I(\vec{y}, x, v) \, Y^I(\Omega_5) \,,
\end{eqnarray}
where the $Y^I$ are scalar $SO(6)$ spherical harmonics and $I = \{ k,
l, m, l', m' \}$ is a total index for the five quantum numbers
characterizing an element of an $SO(6)$ representation.  The label $k$
gives the total $SO(6)$ representation as the $k$-fold symmetric
traceless product of the ${\bf 6}$, while $\{ l,m \}$ and $\{ l',m'
\}$ are the quantum numbers for the $SU(2)_H \times SU(2)_V$ subgroup.
These spherical harmonics are discussed in the appendix, where we show
that the only harmonics that are nonvanishing on the D5-brane ($\psi =
0$) are those with $l' = m' = 0$.  Hence the closed-string modes that
participate in the interactions (\ref{OnePoint}), (\ref{TwoPoint}) and
(\ref{MixedPoint}) are characterized only by $k$, $l$ and $m$.
Furthermore, at $\psi = 0$ the functional form of the harmonic does
not depend on $k$; the total quantum number only determines an overall
normalization.

Let us now consider the one-point couplings (\ref{OnePoint}).  For the dilaton we find
\begin{eqnarray}
S_{\Phi}^{(1)} &=& - \tf12 \, T_{D5} L^6 {1 \over \sqrt{4 \pi}} \int d^4x \sqrt{\bar{g}_4} \, d \Omega
\sum_{k,l,m}\, \Phi^k_{lm} (\vec{y}, v) \, Y^l_m(\Omega)\, Z^k_{l,0}(0)  \,, \\
&=& - \tf12 \, T_{D5} L^6 \int d^4x \sqrt{\bar{g}_4} \sum_{k \: {\rm even}}  z(k) \, 
\Phi^k_{00} (\vec{y}, v) \,.
\nonumber
\end{eqnarray}
Here $z(k)\equiv Z^k_{00}(0)$ is a $k$-dependent normalization factor.
We have integrated over the $S^2$, which gives zero for all $Y^l_m$
except the constant mode $Y^0_0 = 1 / \sqrt{4 \pi}$.  We note that
only the representations of $SO(6)$ with $k$ even contain $SU(2)_H
\times SU(2)_V$ singlets; this can be seen by recalling that ${\bf 6}
\rightarrow ({\bf 3}, {\bf 1}) \oplus ({\bf 1}, {\bf 3})$, and hence
by the usual rules for addition of angular momentum, the $SO(6)$
representation with $k$ even or odd only contains $SO(2)_H \times
SU(2)_V$ representations with total spin $l + l'$ even or odd,
respectively.  The remaining closed string modes involve a similar
reduction of vector and tensor spherical harmonics, which we leave for
the future.

\setcounter{equation}{0}
\section{Field theory action}
\label{FieldSec}

We now determine the action for the dual quantum field theory.  In the
absence of the defect, the theory is simply ${\cal N}=4$ Super
Yang-Mills with gauge group $SU(N)$ in four dimensions; this
completely specifies the four-dimensional fields and their bulk
couplings.  We also know that the defect, which breaks the total
supersymmetry to eight supercharges, hosts a three-dimensional
hypermultiplet, which transforms as a fundamental of the bulk gauge
group (see, for example, \cite{HW}).  In principle, the defect action
can be derived as the $\alpha' \rightarrow 0$ limit of the D3/D5-brane
intersection.  However, we will be able to use gauge invariance and
the preserved supersymmetry and R-symmetry to completely determine the
action, given the inputs above.

The preserved spacetime symmetries of the configuration are
three-dimensional translations and Lorentz transformations, as well as
three-dimensional ${\cal N}=4$ supersymmetry, which admits an $SO(4)$
R-symmetry, realized in our case as $SU(2)_V \times SU(2)_H$.  The
gravity dual predicts that the field theory is additionally
superconformal, but these extra symmetries will not be used to
construct the action.  Classical scale invariance will nonetheless be
manifest, with the dimensionless 4D Yang-Mills coupling $g$ the only
parameter.  Whether conformal symmetry persists on the quantum level
is an important test of the correspondence, for which we provide
partial results in section~\ref{ConformalSec}; further results can be
found in \cite{EGK}.

The interactions on the defect involve both 4D and 3D fields.  These
must be coupled in a supersymmetric way, and consequently, one must
develop a procedure for breaking up 4D supermultiplets into sets of
fields that, when restricted to the defect, transform like 3D
supermultiplets.  The method we use is based on work of Hori
\cite{Hori}, who addressed similar questions of defining
supersymmetric interactions on a codimension one hypersurface (in his
case in two dimensions); similar techniques have been employed
previously to effect ordinary dimensional reduction \cite{Gates}.
This method employs superspace: four-dimensional ${\cal N}=1$
superfields $\Upsilon(\vec{y}, x, \theta)$ can be made into
three-dimensional ${\cal N}=1$ superfields $\Upsilon_{3d}(\vec{y},
\Theta)$ by restricting them to the ``superspace boundary'', which
means imposing $x=0$ as well as two linear relations on the four
fermionic coordinates $\theta$.  Invariant three-dimensional actions
involving $\Upsilon_{3d}(\vec{y},\Theta)$ along with inherently
three-dimensional superfields $Q(\vec{y},\Theta)$ can then easily be
constructed.  Such actions possess terms with derivatives transverse
to the defect and hence are not equivalent to actions obtained by direct
dimensional reduction.  In the next subsection we detail the
superspace boundary method in ${\cal N}=1$ superspace.  In the section
that follows, we construct the action for our eight-supercharge field
theory with defect, and discuss the realization of the extended
supersymmetry.

\subsection{The superspace boundary}

We briefly review some elementary facts about superspace, and in the
process fix our notation.  4D ${\cal N}=1$ superspace consists of the
usual bosonic coordinates ($\vec{y}, x$) with $\vec{y}$ a 3-vector as
well as anticommuting coordinates $\theta$.  To facilitate reduction
to three dimensions, our 4D superspace conventions are in a Majorana
form, and hence $\theta$ is a four-component Majorana spinor.
Superfields $\Upsilon(\vec{y},x,\theta)$ are defined on superspace,
and can be expanded in a terminating power series in $\theta$, where
the coefficients $B(\vec{y},x)$ and $F(\vec{y},x)$ are just the
ordinary bosonic and fermionic fields that make up a given
supersymmetry multiplet.  One defines the superspace covariant
derivative $D$ and supersymmetry generator $S$,
\begin{eqnarray}
D &\equiv& {\partial \over \partial \bar{\theta}} + i \gamma^\mu
\theta \partial_\mu \,, \quad \quad S \equiv {\partial \over \partial
\bar{\theta}} - i \gamma^\mu \theta \partial_\mu \,,\\
\{ D_\alpha, \bar{D}_\beta \} &=& - 2 i \gamma^\mu \partial_\mu \,, \quad \quad
\{ S_\alpha, \bar{S}_\beta \} =  2 i \gamma^\mu \partial_\mu \,, \quad \quad
\{ D_\alpha, \bar{S}_\beta \} = 0 \,.
\end{eqnarray}
and the supersymmetry transformation of a superfield $\Upsilon(\vec{y},x,\theta)$
is simply
\begin{eqnarray}
\label{SuperTransform}
\delta \Upsilon(\vec{y},x,\theta)  = (\bar{\eta} S) \Upsilon(\vec{y},x,\theta) \,,
\end{eqnarray}
with $\eta$ Majorana.  The power of superspace lies in the fact that
products of superfields and their covariant derivatives are again
superfields with the transformation law (\ref{SuperTransform}).  By
integrating such products over superspace, one obtains Lagrangians
that are invariant under supersymmetry by construction.  This is often
far more convenient than fashioning a component action term-by-term
and verifying supersymmetry explicitly.

Chiral (antichiral) superfields $\Phi$ ($\bar\Phi$) obey the condition
$RD\Phi = 0$ ($LD\bar\Phi =0$).  We can write
\begin{eqnarray}
\Phi(\vec{y},x,\theta) &=& e^{-{i \over 2} \bar\theta \gamma^\mu \gamma \theta \partial_\mu } \left( \phi(\vec{y},x) + \sqrt{2} \bar\theta L \psi(\vec{y},x) + \bar\theta L \theta F(\vec{y},x) \right) \,, \\
\bar\Phi(\vec{y},x,\theta) &=& e^{+{i \over 2} \bar\theta \gamma^\mu \gamma \theta \partial_\mu} \left( \bar\phi(\vec{y},x) + \sqrt{2} \bar\theta R \psi(\vec{y},x) + \bar\theta R \theta \bar{F}(\vec{y},x) \right) \,, 
\end{eqnarray}
with $\phi$ and $F$ complex scalars and $\psi$ a Majorana spinor.  The
vector superfield $V^a(\vec{y},x,\theta)$ is a real superfield, which
in Wess-Zumino gauge reads
\begin{eqnarray}
\label{VectorSuperfield}
V^a = - \tf12 \bar\theta \gamma^\mu \gamma \theta A^a_\mu + i
(\bar\theta L \theta) (\bar\theta R \lambda^a) - i (\bar\theta R
\theta) (\bar\theta L \lambda^a) - \tf12 (\bar\theta L \theta)
(\bar\theta R \theta) D^a \,,
\end{eqnarray}
while the field strength superfield is
\begin{eqnarray}
(L W)^a_{\alpha} &\equiv& - \tf18 (\bar{D} R D) e^{-2 V^a T^a} (LD)_{\alpha}
 e^{2 V^a T^a} \\
&=& e^{-{i \over 2} \bar\theta \gamma^\mu \gamma \theta \partial_\mu }
\left( -i (L\lambda^a)_\alpha - D^a (L \theta)_\alpha + \tf{i}{2} (L
\gamma^{\mu\nu} \theta)_\alpha F^a_{\mu\nu} + (L \gamma^\mu D_\mu
\lambda)_\alpha (\bar\theta L \theta) \right) \,.  \nonumber
\end{eqnarray}
We define the superspace measures
\begin{eqnarray}
d^2 \theta_L \equiv d \bar\theta L d\theta \,, \quad \quad
d^2 \theta_R \equiv d \bar\theta R d\theta \,, \quad \quad d^4 \theta \equiv d^2 \theta_L d^2 \, \theta_R \,.
\end{eqnarray}
We then have the action integrals
\begin{eqnarray}
\nonumber
 \int d^4x d^4 \theta \, \bar\Phi e^{2V\cdot T} \Phi &=& \int d^4x [
(D_\mu \phi)^\dagger D^\mu \phi - \tf{i}{2} \bar\psi \gamma^\mu
D_\mu \psi + \bar{F} F + \\
&& i \sqrt{2} ( \bar\phi \bar\lambda^a T^a L
\psi - \bar\psi R \lambda^a T^a \phi) - \bar\phi D \phi ] \,, \\ \nonumber
\int d^4x {\tf12} {\rm Im} \int d^2 \theta_R \, \tau (\bar{W} LW)
&=& \int d^4x [ \tf{1}{g^2} \left( - \tf{1}{4} F^a_{\mu\nu}
F^{a \mu\nu} - \tf{i}{2} \bar\lambda^a \gamma^\mu D_\mu \lambda^a +
\tf12 D^a D^a \right) \\ &+& \tf{\theta}{32 \pi^2} F^a_{\mu\nu}
\tilde{F}^{a \mu\nu} ] \,, \nonumber \\ \int d^4x d^2 \theta_R \,
W(\Phi_i) &=& \int d^4x \left( F^i \partial_i W(\phi) - \tf12
(\partial_i \partial_j W(\phi)) \, \bar\psi_i L \psi_j \right) \,,
\nonumber
\end{eqnarray}
with the definitions
\begin{eqnarray}
D_\mu \phi = (\partial_\mu - i A^a_\mu T^a) \phi \,, \quad \quad
D_\mu \psi = (\partial_\mu - i A^a_\mu (L T^a - R T^{*a})) \psi \,, \quad \quad
\tau \equiv {i \over g^2} + {\theta \over 8 \pi^2} \,.
\end{eqnarray}

It is clear that the presence of the defect must break some
supersymmetry, since $x$-translations are broken; supercharges that
anticommute to such translations must also be broken.  The only
possibility is that half the supersymmetry is preserved, leaving 3D
${\cal N}=1$.

Under the three-dimensional Lorentz group, a four-component spinor
decomposes into a pair of two-component 3D spinors, labeled by an
additional index $i= 1,2$.  The decomposition of gamma matrices in our
basis is given in Appendix B.  For example, the four-component
supersymmetry generator $S$ turns into a pair of two-component
objects:
\begin{eqnarray}
\label{SDecomp}
S_1 = {\partial \over \partial \bar{\theta}^1} - i \rho^k \theta_1 \partial_k
+ \theta_2 \partial_x \,, \quad \quad
S_2 = - {\partial \over \partial \bar{\theta}^2} + \rho^k \theta_2 \partial_k
+ \theta_1 \partial_x \,.
\end{eqnarray}
Only a linear combination of the generators (\ref{SDecomp}) that does
not involve $\partial_x$ can be preserved.  To this end, we must place
two linear relations on the four $\theta$ coordinates: a convenient
choice for us is
\begin{eqnarray}
\theta_2 = 0 \,,
\end{eqnarray}
where we bear in mind $\theta_2$ is a two-component real 3D spinor.
Defining $\Theta \equiv \theta_1$, we now have the 3D ${\cal N}=1$
superspace covariant derivative and supersymmetry generator
\begin{eqnarray}
{\cal D} \equiv D_1 |_{\theta_2 = 0} = {\partial \over \partial \bar{\Theta}} + i \rho^k \Theta \partial_k \,, \quad \quad
{\cal S} \equiv S_1 |_{\theta_2 = 0} = {\partial \over \partial \bar{\Theta}} - i \rho^k \Theta \partial_k \,. 
\end{eqnarray}
Fields native to the defect are naturally written as inherently 3D
superfields $Q(\vec{y}, \Theta)$.  These have the expansion
\begin{eqnarray}
Q(\vec{y},\Theta) = q(\vec{y}) + \bar\Theta \Psi(\vec{y}) + \tf12 \bar\Theta \Theta f(\vec{y}) \,, 
\end{eqnarray}
and may be real or complex, but if complex, the real and imaginary
parts transform independently under supersymmetry.  
Furthermore, from any 4D superfield $\Upsilon(\vec{y},x,\theta)$ we may
create a 3D superfield $\Upsilon_{3d}(\vec{y},\Theta)$ by restricting to the
``superspace boundary'':
\begin{eqnarray}
\Upsilon_{3d}(\vec{y}, \Theta) = \Upsilon(\vec{y},x,\theta)|_\partial
 \equiv \Upsilon(\vec{y},x,\theta)|_{x=\theta_2=0} \,.
\end{eqnarray}
This is the central concept.  $\Upsilon_{3d}(\vec{y},\Theta)$ includes
some or all of the component fields contained in
$\Upsilon(\vec{y},x,\theta)$ restricted to the defect at $x=0$.  As
can readily be seen, $\Upsilon_{3d}$ transforms as a 3D superfield
under the preserved supersymmetry transformations, namely
(\ref{SuperTransform}) with $\eta_2=0$.  Consequently, any product
of $\Upsilon_{3d}(\vec{y},\Theta)$ and $Q(\vec{y},\Theta)$ and their
3D covariant derivatives $(Q^i(\vec{y}, \Theta) \cdots
\Upsilon_{3d}^a(\vec{y}, \Theta) \cdots {\cal D}Q^j(\vec{y}, \Theta)
\cdots {\cal D}\Upsilon_{3d}^b(\vec{y}, \Theta) \cdots)$ may be
integrated over the two $\Theta$ coordinates to produce a 3D ${\cal
N}=1$ invariant Lagrangian. We define the measure
\begin{eqnarray}
d^2 \Theta \equiv \tf12 d\bar\Theta d\Theta \,.
\end{eqnarray}
As an example of a 4D superfield restricted to the superspace
boundary, we find for the chiral superfield $\Phi$,
\begin{eqnarray}
\label{ReduceChiral}
\Phi |_\partial = \phi + \frac{1}{\sqrt{2}} \bar\Theta (\psi_1 - i \psi_2) + \tf12 \bar\Theta \Theta ( F + i \partial_3 \phi) \,,
\end{eqnarray}
where again $\psi_1$, $\psi_2$ are the 2-component spinors emerging
from the 4-component $\psi$.  The real and imaginary parts of
(\ref{ReduceChiral}) transform independently under 3D ${\cal N}=1$
supersymmetry, exhibiting the decomposition of a 4D ${\cal N}=1$
chiral multiplet into two 3D ${\cal N}=1$ real multiplets.

The appearance of the transverse derivative $\partial_3 \phi$ in
(\ref{ReduceChiral}) may appear at first unusual, but it is required
by 3D supersymmetry (as one may easily check using component
transformations), and will prove vital in our construction of the
eight-supercharge Lagrangian.  When one compactifies the 3-direction
and expands $\phi$ in normal modes, $\partial_3 \phi$ contributes the
appropriate mass terms to the 3D auxiliary field, which helps in
understanding its presence.

In three dimensions, the superspace action for the kinetic terms of
superfields $Q$ as well as coupling to a gauge multiplet is \cite{GGRS}
\begin{eqnarray}
\label{3DAction}
S_{kin} = \int d^3x d^2\Theta \, \tf12 (\overline{\nabla Q}) \nabla Q \,,
\end{eqnarray}
where we have defined the superspace gauge covariant derivative
\begin{eqnarray}
\label{Nabla}
\nabla \equiv {\cal D} - i \Gamma^a T^a \,,
\end{eqnarray}
including the connection spinor superfield $\Gamma^a$, which
contains the gauge field and its partners.  We are not interested in
inherently 3D gauge fields, but instead we wish to obtain a connection
superfield by starting with some 4D superfield containing the gauge
multiplet and reducing to the superspace boundary.  We arrive
at\footnote{In principle one could define $\Gamma^a T^a \equiv
{\tf{1}{\alpha}} e^{-i \alpha V^a T^a} D e^{i \alpha V^a T^a}$ for
any $\alpha$, but upon setting $\theta_2 = 0$ these all coincide in
Wess-Zumino gauge.  Outside WZ gauge, $\alpha$ would appear in the
coupling of the gauge-artifact fields in the vector multiplet to the
3D modes, but these terms have no physical content.}
\begin{eqnarray}
\label{Connection}
\Gamma^a \equiv (DV^a)_2 |_\partial = i \rho^k \Theta A^a_k +\lambda^a_1 (\bar\Theta \Theta) \,.
\end{eqnarray}
Here we decompose the 4D spinor $D V^a$ into two-component 3D spinors
and keep the latter 3D spinor, restricting it to the superspace
boundary.  Notice that the auxiliary $D^a$ does not survive the
projection to three dimensions; this is appropriate since a 3D ${\cal
N}=1$ vector multiplet does not contain an auxiliary field
\cite{GGRS}. With the definition (\ref{Connection}), the action
(\ref{3DAction}) reduces tgo
\begin{eqnarray}
S_{kin} = \int d^3x \left( (D^k q)^\dagger D_k q - i \bar\Psi \rho^k D_k \Psi + \bar{f} f + i \bar{q} \bar\lambda^a_1 T^a \Psi - i \bar\Psi \lambda^a_1 T^a q \right)\,,
\end{eqnarray}
with $D_k = \partial_k - i T^a A^a_k$.
This indeed contains a canonical coupling between 3D matter, and
(certain components of) the 4D gauge field and its superpartners.  In
the next subsection, we will apply these results to obtain the
particular ${\cal N}=4$ theory we need to describe our system.

Besides making supersymmetry manifest, the ``superspace boundary''
technique outlined here has the advantage of producing an action
already formulated in superspace language.  This facilitates
perturbation theory, where all but the most elementary calculations in
component formalism prove far too cumbersome even in the case of the
pure bulk ${\cal N}=4$ SYM.

A drawback of using this superspace formalism for our system, however,
is that it makes only one quarter of the supersymmetry manifest: four
supercharges in the bulk broken to two on the defect, instead of
sixteen broken to eight.  This also means that the R-symmetries are
obscured: only a diagonal $SU(2)_D \subset SU(2)_V \times SU(2)_H$
will be visible in the superspace action.  To confirm that the larger
symmetries are present, we will reduce to a component action, and
explicitly demonstrate $SU(2)_V \times SU(2)_H$ invariance.  The
existence of this R-symmetry then implies the full 3D ${\cal N}=4$
supersymmetry.

\subsection{Action for field theory with defect}
\label{ActionSec}

Under the reduced supersymmetry, the bulk 4D ${\cal N}=4$ vector
multiplet decomposes into a 3D ${\cal N}=4$ vector multiplet and a 3D
${\cal N}=4$ adjoint hypermultiplet.  As described in \cite{HW}, the
bosonic components of the vector multiplet are $\{ A_k, X^7, X^8,
X^9 \}$, with the scalars transforming as the ${\bf 3}$ of $SU(2)_V$,
while the hypermultiplet consists of $\{ A_6, X^3, X^4, X^5 \}$, with
these scalars a triplet of $SU(2)_H$.  (In fact, we will see this is
slightly oversimplified: the $x$-derivatives of $X^3$, $X^4$ and $X^5$
actually participate in the vector multiplet, as does $A_6$,
as part of the auxiliary field.)  The four adjoint Majorana spinors of
${\cal N}=4$ SYM transform as a $({\bf 2}, {\bf 2})$ of $SU(2)_V
\times SU(2)_H$, which we denote $\lambda^{im}$. Under the reduced
spacetime symmetries, they decompose into pairs of two-component 3D
Majorana spinors, with $\lambda_1^{im}$ ending up in the vector
multiplet and $\lambda_2^{im}$ in the hyper.

The hypermultiplet living on the defect transforms in the fundamental
representation of the gauge group.  It consists of an
$SU(2)_H$-doublet of complex scalars $q^m$ and an $SU(2)_V$-doublet of
Dirac 3D fermions $\Psi^i$.  In addition to the R-symmetry charges,
the defect hypermultiplet is also charged under a global $U(1)_B$,
under which the bulk fields are inert; the corresponding current is
dual to the D5-brane gauge field on the gravity side.  Because the
defect hyper fields are in the fundamental representation of the gauge
group $SU(N)$, they are coupled canonically to $A_k$, and hence
supersymmetry will induce couplings to the rest of the bulk vector
multiplet as well, which we determine below.  The bulk hypermultiplet
does not directly couple to the defect fields.

The field content and Lagrangian for the theory in the bulk are
identical to that of ${\cal N}=4$ Super Yang-Mills with gauge group
$SU(N)$.  Using ${\cal N}=1$ superspace, the superfields are an
$SU(N)$ vector multiplet $V^a$, as in (\ref{VectorSuperfield}), and
three chiral multiplets in the adjoint representation, $X^{aA}$ with
$A=1,2,3$:
\begin{eqnarray}
\Phi^{aA} &=& e^{-{i \over 2} \bar\theta \gamma^\mu \gamma \theta \partial_\mu } \left( X^{aA} + \sqrt{2} \bar\theta L \chi^{aA} + \bar\theta L \theta F^{aA} \right) \,,
\end{eqnarray}
and the ${\cal N}=4$ action in our conventions is
\begin{eqnarray}
\nonumber
S_4 &=& S_K + S_g + S_W \,, \\
\label{S4}
S_K &=& \tf{1}{g^2} \int d^4x \, d^4 \theta \, \bar\Phi^{Ab}
\left(e^{2 V^a t^a}\right)_{bc} \Phi^{Ac} \,, \\ \nonumber S_g &=&
\int d^4x \, {\tf12} \, {\rm Im} \int d^2 \theta_R \, \tau (\bar{W}
LW) \,, \\ \nonumber S_W &=& \tf{1}{g^2} \int d^4x \epsilon_{ABC}
f^{abc} \tf{\sqrt{2}}{3!} \left( \int d^2 \theta_R \, \Phi^{Aa}
\Phi^{Bb} \Phi^{Cc} + \int d^2 \theta_L \, \bar{\Phi}^{Aa}
\bar{\Phi}^{Bb} \bar{\Phi}^{Cc} \right) \,,
\end{eqnarray}
where $(t^a)_{bc} = - i f^{abc}$ since the $\Phi^{Aa}$ are in the adjoint
representation.  In components, this is
\begin{eqnarray}
\nonumber S_4 &=& \tf{1}{g^2} \int d^4x [ - \tf{1}{4}
F^a_{\mu\nu} F^{a \mu\nu} - \tf{i}{2} \bar\lambda^a \gamma^\mu D_\mu
\lambda^a + \tf12 D^a D^a + \tf{\theta}{32 \pi^2} F^a_{\mu\nu}
\tilde{F}^{a \mu\nu} \\ \label{AmbientComps} &+& (D^\mu
X^{Aa})^\dagger D_\mu X^{Aa} - \tf{i}{2} \bar\chi^{Aa} \gamma^\mu
D_\mu \chi^{Aa} + F^{Aa} \bar{F}^{Aa} \\ &+& \sqrt{2} f^{abc}
(\bar{X}^{Ab} \bar\lambda^a L \chi^{Ac} - \bar\chi^{Ab} R \lambda^a
X^{Ac}) + i f^{abc} \bar{X}^{Ab} D^a X^{Ac} \nonumber \\ &+& 
\tf{1}{\sqrt{2}} \, \epsilon_{ABC} f^{abc} (F^{Aa} X^{Bb} X^{Cc} +
\bar{F}^{Aa} \bar{X}^{Bb} \bar{X}^{Cc} - \bar\chi^{Aa} (L X^{Cc} + R
\bar{X}^{Cc}) \chi^{Bb})] \,, \nonumber 
\end{eqnarray}
with $D_\mu X^a = \partial_\mu X^a + f^{abc} A_\mu^b X^c$ and likewise
for the fermions.

The defect hypermultiplet can be written as two complex 3D multiplets
$Q^i$, $i=1,2$:
\begin{eqnarray}
Q^i &=& q^i + \bar\Theta \Psi^i + \tf12 \bar\Theta \Theta f^i \,, \\
\bar{Q}^i &=& \bar{q}^i + \bar\Psi^i \Theta + \tf12 \bar\Theta \Theta
\bar{f}^i \,.
\nonumber
\end{eqnarray}
The superfields $Q^i$ ($\bar{Q}^i$) transform in the fundamental
(antifundamental) representation of $SU(N)$; we have suppressed the
gauge indices.  They are coupled to the bulk gauge fields in the way
we have outlined:
\begin{eqnarray}
\label{ThreeKinetic}
S_{kin} = \tf{1}{g^2} \int d^3x d^2\Theta \, \tf12 (\overline{\nabla Q^i})
\nabla Q^i \,,
\end{eqnarray}
with $\nabla$ as in (\ref{Nabla}).

Finally, to obtain a theory that preserves 8 supercharges and places
the 3D part of the gauge field $A_k$ in a single supermultiplet with
the scalars $X^7$, $X^8$, $X^9$, we must produce a coupling of the
$Q^i$ to half the fields in the $\Phi^A$.  We choose a convention
where the scalar parts of $\Phi^A$ are $(X_V^A + i X_H^A) / \sqrt{2}$,
with $X_H = (X^3, X^4, X^5)$ and $X_V = (X^7, X^8, X^9)$.  We then
define the following 3D superfields by restricting $\Phi^{Aa}$ to the
superspace boundary:
\begin{eqnarray}
\label{DefineX}
{\cal X}^{Aa} T^a &\equiv& {\rm Re} \,( e^{V \cdot T} \Phi^{Aa} T^a e^{-V\cdot T})|_{\partial} \\  &=& \left( {\rm Re}\, X^{Aa} + {\tf{1}{\sqrt{2}}} \bar\Theta \chi_1^{Aa} + {\tf12} \bar\Theta \Theta ( {\rm Re}\, F^{Aa} - \partial_6 \, {\rm Im}\, X^{Aa} - f^{abc} A_6^b \, {\rm Im}\,X^{Ac}) \right) T^a \,, \nonumber\\
&\equiv& {\tf{1}{\sqrt{2}}} \left( X_V^{Aa} +  \bar\Theta \chi_1^{Aa} + {\tf12} \bar\Theta \Theta (F^{Aa}_V - D_6 X_H^{Aa}) \right) T^a\,,
\nonumber
\end{eqnarray}
where $T^a$ are generators in the fundamental representation of
$SU(N)$.  The sole consequence of the exponential terms in the
definition (\ref{DefineX}) is to covariantize the transverse
derivative $\partial_6$, which is necessary to preserve 4D gauge
invariance.  We now claim that the final piece of the action is
\begin{eqnarray}
\label{ThreePotential}
S_{X} = \tf{1}{g^2} \int d^3x d^2\Theta \, \sqrt{2}\, \sigma^A_{ij}
\,\bar{Q}^i {\cal X}^{Aa} T^a Q^j \,,
\end{eqnarray}
where $\sigma^A_{ij}$ are the Pauli matrices.  This is the ${\cal
N}=4$ supersymmetric completion of (\ref{ThreeKinetic}), and therefore
involves the same coupling constant, $g$.  Hence the defect action
adds no new couplings to the theory.  That (\ref{ThreePotential}) is
bilinear in $Q^i$ and linear in the ${\cal X^A}$ can be expected on
the grounds of gauge invariance and supersymmetry.  The origin of the
precise coefficients will emerge as we discuss the symmetries and
component expansion of this action.

We notice immediately that, not only the scalars $X^7$, $X^8$ and
$X^9$, but also the fields $X^3$, $X^4$, $X^5$ and $A_6$ participate
in the bulk vector multiplet and couple to the boundary
hypermultiplet, due to the $D_6 X_H^{Aa}$ term inside the auxiliary
field of ${\cal X}^{Aa}$.  This should not be too surprising, since it
is known that constraining the bulk vector multiplet to vanish at the
defect places Dirichlet boundary conditions on $X_V^A$ and Neumann
boundary conditions on $X_H^A$ \cite{HW}.  Analogously, the bulk
hypermultiplet restricted to the defect contains the first derivatives
of the $X_V^A$ along with the restriction of the $X_H^A$.

Let us examine how the symmetries of the system are realized in the
action (\ref{S4}), (\ref{ThreeKinetic}), (\ref{ThreePotential}).
${\cal N}=4$ SYM has an $SU(4)_R$ R-symmetry, of which only $SU(3)
\times U(1)_R$ is visible in the ${\cal N}=1$ superspace formulation:
the $SU(3)$ acts on the three chiral superfields in the obvious way.
Once the defect is introduced, only $SU(2)_V \times SU(2)_H \subset
SU(4)_R$ is preserved.  We cannot hope that more than the intersection
of $SU(2)_V \times SU(2)_H$ with $SU(3) \times U(1)_R$ will be visible
in our presentation.  In our convention for the components of $\Phi$,
the $SO(3) \subset SU(3)$ is precisely the diagonal subgroup $SU(2)_D
\subset SU(2)_V \times SU(2)_H$, and this turns out to be the manifest
part of the R-symmetry.  Under $SU(2)_D$, the defect hyper fields
$q^i$ and $\Psi^i$ should both transform as a doublet.  Hence
$SU(2)_D$ acts as a global symmetry on our superfields: $Q^i$ is a
doublet and ${\cal X}^A$ is a triplet, while $\Gamma$ is a singlet.
The kinetic action (\ref{ThreeKinetic}) is obviously invariant under
$SU(2)_D$; preserving the symmetry in (\ref{ThreePotential}) requires
the Pauli matrix coupling, but does not specify the overall
coefficient.

The global $U(1)_B$ symmetry, with current dual to the D5-brane gauge
field, is also manifest in the superspace presentation: the superfield
$Q^i$ has charge one while the bulk fields are inert.

Also worth mentioning are a pair of discrete parity symmetries, $P$
and $P_6$.  In three dimensions, reversing the sign of both spatial
coordinates is a part of the proper Lorentz group, but reversing the
sign of just one, which we call $P$, is nontrivial.  For example, we
can send $x_2 \rightarrow - x_2$, $A_2 \rightarrow - A_2$.  The total
bulk and defect superspace action (\ref{S4}), (\ref{ThreeKinetic}),
(\ref{ThreePotential}) is then invariant\footnote{Assuming the
vanishing of the vacuum $\theta$-angle.} under the transformation
\begin{eqnarray}
\label{Parity}
P: \quad \quad \theta \rightarrow i \gamma^2 \theta \,, \quad \quad V^a \rightarrow -V^a \,, \quad \quad \Phi \rightarrow - \bar{\Phi} \,, \quad \quad Q \rightarrow Q \,.
\end{eqnarray}
One can also consider reversing the sign in the broken direction, $x_6
\rightarrow - x_6$, $A_6 \rightarrow - A_6$.  This is realized on superspace as
\begin{eqnarray}
\label{Parity6}
P_6: \quad \quad \theta \rightarrow i \gamma \gamma^3 \theta \,, \quad \quad V^a \rightarrow -V^a \,, \quad \quad \Phi \rightarrow \bar{\Phi} \,, \quad \quad Q \rightarrow Q \,.
\end{eqnarray}
The superspace transformations (\ref{Parity}) and (\ref{Parity6})
implicitly determine the action of parity on the component fields.
The transformation $P_6$ is realized trivially on our defect action,
as it is equivalent to changing the signs of the ambient hyper fields
(which do not participate) while leaving the ambient vector and defect
fields inert.  It is a nontrivial symmetry of the ${\cal N}=4$ SYM action.  

Not evident in the superspace formulation are the remaining
off-diagonal symmetries in $SU(2)_V \times SU(2)_H$.  Under a
$SU(2)_V$ transformation, $\Psi^i$ will rotate while $q^i$ is inert,
and the converse for $SU(2)_H$.  Additionally, under a generic
$SU(2)_V \times SU(2)_H$ transformation, the fermi fields $\chi_1^{Aa}$
inside ${\cal X}^{Aa}$ mix with the ${\cal N}=1$ gaugino $\lambda^a_1$
inside $\Gamma^a$, and together form a $({\bf 2}, {\bf 2})$.  It is
obvious that if these symmetries are present, they will only be
visible by reducing to the component action.

In components, the defect action (\ref{ThreeKinetic}), (\ref{ThreePotential}) is
\begin{eqnarray}
\label{S3}
S_3 &=& S_{kin} + S_{X} \,, \\
\label{Skin}
S_{kin} &=& \tf{1}{g^2} \int d^3x \left( (D^k q^i)^\dagger D_k q^i - i \bar\Psi^i \rho^k D_k \Psi^i + \bar{f}^i f^i + i \bar{q}^i \bar\lambda^a_1 T^a \Psi^i - i \bar\Psi^i \lambda^a_1 T^a q^i \right)\,, \\
S_{X} &=& \tf{1}{g^2}  \int d^3x \, [ -  \sigma^A_{ij} \bar\Psi^i X_V^{Aa} T^a \Psi^j - \sigma^A_{ij} ( \bar{q}^i \bar\chi_1^{Aa} T^a \Psi^j + \bar\Psi^i \chi_1^{Aa} T^a q^j) \label{SX} \\
&& + \sigma^A_{ij} ( \bar{q}^i X_V^{Aa} T^a f^j + \bar{f}^i X_V^{Aa} T^a q^j + \bar{q}^i ( F_V^{Aa} - D_6 \, X_H^{Aa}) T^a q^j)]\,. \nonumber
\end{eqnarray}
We would like to demonstrate the full $SU(2)_V \times SU(2)_H$
invariance.  The kinetic terms are obviously invariant.  Let us next
examine the Yukawa terms coupling the defect hyper to the bulk
fermions $\lambda_1$, $\chi_1^A$.  We define the gaugino fields
\begin{eqnarray}
\lambda^a_{im} \equiv \lambda^a \delta_{im} - i \chi^{Aa} \sigma^A_{im} \,, 
\end{eqnarray}
which transform as $\lambda^a \rightarrow g_V \lambda^a
{g_H}^\dagger$, analogous to a linear sigma model field $\sigma + i
\pi^A \sigma^A$.  Here we are using $i$, $j$ as $SU(2)_V$ indices and
$m$, $n$ as $SU(2)_H$ indices.  The Yukawa terms then become
\begin{eqnarray}
\label{FullYukawa}
\int d^3x \, (i \bar{q}^m (\bar{\lambda}^a_1)_{mi} T^a \Psi^i - i
\bar{\Psi}^i (\lambda^a_1)_{im} T^a q^m ) \,,
\end{eqnarray}
and are manifestly invariant.  The precise value of the coefficient in
(\ref{ThreePotential}) was required to construct (\ref{FullYukawa}).
There is one more Yukawa term in (\ref{SX}), namely
\begin{eqnarray}
\label{HyperinoYukawa}
- \int d^3x \, \sigma^A_{ij} \, \bar\Psi^i X_V^{Aa} T^a \Psi^j \,.
\end{eqnarray}
This obviously respects $SU(2)_V \times SU(2)_H$: $X^A_V$ is a triplet
of $SU(2)_V$ and $\Psi^i$ is a doublet, and all fields are inert under
$SU(2)_H$.  Furthermore, the scalar derivative coupling
\begin{eqnarray}
- \int d^3x \, \sigma^A_{ij} \, \bar{q}^i (D_6 \, X_H^{Aa}) \, T^a q^j \,,
\end{eqnarray}
transforms under $SU(2)_H$ in the same way (\ref{HyperinoYukawa})
did under $SU(2)_V$, and is similarly invariant.

Finally we come to the auxiliary fields and the scalar potential.
Having entirely fixed the form of (\ref{ThreePotential}) to enforce
$SU(2)_V \times SU(2)_H$ on the Yukawa terms, invariance in this
sector is a nontrivial check, and in fact we find a gratifying
interplay between bulk and defect auxiliary fields that preserves the
symmetries.  The result is reminiscent of how in the bulk ${\cal N}=4$
SYM theory, neither $F$-term nor $D$-term contributions to the scalar
potential are individually $SU(4)$ invariant, but instead only the
sum.

Considering first the defect auxiliaries $f^i$, we have the terms
\begin{eqnarray}
\label{Actionf}
\int d^3x \left( \bar{f}^i f^i +  \sigma^A_{ij} \, ( \bar{q}^i X_V^{Aa} T^a f^j + \bar{f}^i X_V^{Aa} T^a q^j ) \right) \,.
\end{eqnarray}
Eliminating the $f^i$ via their equations of motion as usual, we find
\begin{eqnarray}
\label{fSolve}
f^i = -  \sigma^A_{ij} \, X_V^{Aa} T^a q^j \,, \quad \quad
\bar{f}^j = - \sigma^A_{ij} \, \bar{q}^i X_V^{Aa} T^a \,,
\end{eqnarray}
and then (\ref{Actionf}) becomes
\begin{eqnarray}
\int d^3x \, (- \bar{f}^i f^i) = - \int d^3x \, \bar{q}^a (\sigma^A
\sigma^B)_{ij} T^a T^b q^j \, X_V^{Ab} X_V^{Bb} \,.
\end{eqnarray}
Using the relation $\sigma^A \sigma^B = \delta^{AB} + i \epsilon_{ABC}
\sigma^C$ and symmetrization/antisymmetrization, we obtain the result
\begin{eqnarray}
\label{Potentialf}
\int d^3x \left( - \tf12 \bar{q}^i \{ T^a, T^b \} q^i \, X_V^{Aa} X_V^{Ab}
+ \tf12 \epsilon_{ABC} f^{abc} \, \bar{q}^i \, \sigma_{ij}^A \, T^a q^j \, X_V^{Bb} X_V^{Cc} \right)\,.
\end{eqnarray}
The first term is $SU(2)_V \times SU(2)_H$ invariant, since the $q$
and $X_V$ variations cancel separately.  The second term, however, is not
invariant.  Fortunately, we have not exhausted the contributions to
the potential.

We turn now to the bulk auxiliary fields.  Their action can be written
\begin{eqnarray}
\int d^4x \left( \bar{F}^{Aa} F^{Aa} + {\tf{1}{\sqrt{2}}}
\epsilon_{ABC} f^{abc} ( F^{Aa} X^{Bb} X^{Cc} + \bar{F}^{Aa} \bar{X}^{Bb}
\bar{X}^{Cc}) + \delta(x_6) \, \bar{q}^i \sigma^A_{ij} T^a q^j F_V^{Aa}  \right) \,,
\end{eqnarray}
where the last term comes from the defect action. In terms of real and imaginary parts, this becomes
\begin{eqnarray}
\int d^4x [ \tf12 (F_V^{Aa} F_V^{Aa} + F_H^{Aa} F_H^{Aa} &+& 
\epsilon_{ABC} f^{abc} (F_V^{Aa} X_V^{Bb} X_V^{Cc} - F_V^{Aa} X_H^{Bb}
X_H^{Cc} - 2 F_H^{Aa} X_H^{Bb} X_V^{Cc})) \nonumber \\ &+& \delta(x_6)
 \, \bar{q}^i \sigma^A_{ij} T^a q^j F_V^{Aa}] \,.
\end{eqnarray}
The imaginary part $F_H$ does not couple to the defect, so its
contribution to the potential is unchanged from ${\cal N}=4$ SYM.
For the real part $F_V$, we find
\begin{eqnarray}
F_V^{Aa} = -  \left( \tf12 \epsilon_{ABC} f^{abc} ( X_V^{Bb} X_V^{Cc} - X_H^{Bb} X_H^{Cc} ) + \delta(x_6) \, \bar{q}^i \sigma^A_{ij} T^a q^j \right) \,,
\end{eqnarray}
where the first part is the same as ${\cal N}=4$ SYM.  Thus all terms from
the bulk auxiliaries $F^A$ are
\begin{eqnarray}
&&\int d^4x - \tf12 \left( F_V^{Aa} F_V^{Aa} + F_H^{Aa} F_H^{Aa} \right) = \\
&&\int d^4x \left( - V_4^F - \tf12 \, \delta(x_6) \, \epsilon_{ABC} f^{abc} ( X_V^{Bb} X_V^{Cc} - X_H^{Bb} X_H^{Cc} ) \,\bar{q}^i \sigma^A_{ij} T^a q^j - \tf12 \delta(x_6)^2 \, (\bar{q}^i \sigma^A_{ij} T^a q^j)^2 \right) \,.\nonumber
\end{eqnarray}
Here $V_4^F$ is the usual $F$-term contribution to the ${\cal N}=4$ SYM
potential, which when combined with the bulk $D$-terms is of course
$SU(2)_V \times SU(2)_H$ invariant (in fact it's $SU(4)$ invariant).
The $\delta(x_6)^2$ term is also obviously invariant.  The remaining
terms can be integrated over $\delta(x_6)$ to produce a
three-dimensional potential.  ``Miraculously'', the $X_V^{Bb}
X_V^{Cc}$ term exactly cancels the non-invariant piece from
(\ref{Potentialf}).  The final term is invariant, as both $\bar{q}
\sigma^A q$ and $\epsilon_{ABC} X_H^B X_H^C$ are triplets of $SU(2)_H$
and singlets of $SU(2)_V$.

We have now demonstrated that besides being 3D ${\cal N}=1$
supersymmetric by construction, our bulk/defect action has an $SU(2)_V
\times SU(2)_H$ R-symmetry.  We therefore conclude that it is in fact
3D ${\cal N}=4$ supersymmetric.  We summarize the final expression for
the defect action, including the potential:
\begin{eqnarray}
S_3 &=& S_{kin} + S_{yuk} + S_{pot} \,, \\
S_{kin} &=& \tf{1}{g^2} \int d^3x \left( (D^k q^m)^\dagger D_k q^m
- i \bar\Psi^i \rho^k D_k \Psi^i \right) \,, \\
S_{yuk} &=& \tf{1}{g^2} \int d^3x \left(i \bar{q}^m
(\bar{\lambda}^a_1)_{mi} T^a \Psi^i - i \bar{\Psi}^i
(\lambda^a_1)_{im} T^a q^m - \bar\Psi^i \sigma^A_{ij} X_V^{Aa} T^a \Psi^j\right)  \,, \\
\label{Spot}
S_{pot} &=& 
- \tf{1}{g^2} \int d^3x \tf12 \left( \bar{q}^m \{ T^a, T^b \} q^m \, X_V^{Aa} X_V^{Ab} -
\epsilon_{IJK} f^{abc} X_H^{Jb} X_H^{Kc}  \,\bar{q}^m \sigma^I_{mn} T^a q^n \right) \\
&& - \tf{1}{g^2} \int d^3x \left(\bar{q}^m \sigma^I_{mn} (D_6 \, X_H^{Ia})\, T^a q^n +  \tf12 \, \delta(0)  (\bar{q}^m \sigma^I_{mn} T^a q^n)^2  \right)
\,, \nonumber 
\end{eqnarray}
where we have distinguished $SU(2)_V$ indices $i,j,A,B,C$ from
$SU(2)_H$ indices $m,n,I,J,K$.  The $\delta(0)$ factor in (\ref{Spot})
may seem curious, but in fact terms of this nature have already been
anticipated by Kapustin and Sethi \cite{KS}, who argued they were
necessary to obtain a sensible Higgs branch, and by Mirabelli and
Peskin \cite{Peskin}, who showed them to be necessary for proper
cancellation of divergences in a 5D case.  Such terms are a generic
feature of supersymmetric couplings of defect matter to
higher-dimensional gauge multiplets involving auxiliary fields.  We
shall have more to say about $\delta(0)$ in
section~\ref{ConformalSec}.

Before leaving the action behind, let us discuss a few other terms
that one might try to include, and argue on symmetry grounds that they
are absent.  In particular, to justify our action we must rule out the
presence of other marginal couplings.  Doing so has the additional
benefit that the gauge coupling $g$ is left as the unique parameter of
the defect theory.  ${\cal N}=1$ supersymmetry does not forbid terms
of the form
\begin{eqnarray}
\label{SQuartic}
S_{quartic} &=& \int d^3x d^2 \Theta \left( \bar{Q}^i Q^i  \bar{Q}^j Q^j \right) 
= \int d^3x \, [2( \bar{f}^i q^i \bar{q}^j q^j + \bar{q}^i f^i \bar{q}^j q^j) \\
&-& (2 \bar{\Psi}^i \Psi^i \bar{q}^j q^j + 2 \bar{\Psi}^i q^i \bar{q}^j \Psi^j - \bar{q}^i \Psi^{Ti} \rho^0 \bar{q}^j \Psi^j - \bar{\Psi}^i q^i \Psi^{\dagger j} q^j )]   \,.
\nonumber
\end{eqnarray}
The two independent ways of contracting the gauge indices lead to two
dimensionless couplings, which generically run with scale.
Eliminating the $f$ fields results in the new contributions to the
scalar potential
\begin{eqnarray}
\label{QuarticTerms}
(\bar{q}^i q^i)^3 \,, \quad \quad (\bar{q}^i q^i) \sigma^A_{jk} \bar{q}^j X_V^{Aa} T^a q^k \,. 
\end{eqnarray}
The $SU(2)_V \times SU(2)_H$ R-symmetry of our theory, however, does
not permit us to modify the action with (\ref{SQuartic}); although the
$(\bar{\Psi}^i \Psi^i \bar{q}^j q^j)$ and $(\bar{q}^i q^i)^3$ terms
are $SU(2)_V \times SU(2)_H$-invariant, the rest are not.

We have assumed throughout this section that the defect couples only
to the bulk vector multiplet, and that the bulk hypermultiplet ignores
the localized matter at tree level.  One can readily see that a term
analogous to (\ref{ThreePotential}) but involving 
\begin{eqnarray}
{\cal Y}^{Aa} T^a \equiv {\rm Im} \,( e^{V \cdot T} \Phi^{Aa} T^a e^{-V\cdot T})|_{\partial} = {\tf{1}{\sqrt{2}}}  \left( X_H^{Aa} - \bar\Theta \chi_2^{Aa} + {\tf12} \bar\Theta \Theta (F^{Aa}_H + D_6 X_V^{Aa}) \right) T^a\,,
\end{eqnarray}
instead of ${\cal X}^{Aa}$ is forbidden, since the $SU(2)_V \times
SU(2)_H$ assignments of the participating bulk scalars are reversed.
(Such a term would be part of the mirror coupling of the defect matter
to the bulk hyper only, wherein the $SU(2)_V \times SU(2)_H$ charges
of $q^m$ and $\Psi^i$ are exchanged.)

One may also consider interactions on the defect involving only the
ambient fields.  The marginal term
\begin{eqnarray}
\label{CS}
S_{CS} = \tf{1}{2g^2}\int d^3x d^2 \Theta  \, \left( \bar{\Gamma}^\alpha (\bar{D}^\beta D_\alpha \Gamma_\beta) + \cdots \right)\,,
\end{eqnarray}
leads to both a gaugino bilinear $(\bar{\lambda}^a_1 \lambda^a_1)$ and
a Chern-Simons term $(\epsilon^{klm} A^a_k \partial_l A^a_m + \cdots)$
for the restriction of the gauge field; the ellipsis in (\ref{CS})
indicates terms with 3 and 4 factors of $\Gamma$ necessary for the
non-Abelian completion \cite{GGRS}.  Notice that while fermion
bilinears and a Chern-Simons piece for inherently three-dimensional
fields would be mass terms, for the ambient fields localized on the
brane they are marginal.  Such terms are related by ${\cal N}=4$
supersymmetry to
\begin{eqnarray}
\label{Xmasses}
S_{X^2} = \int d^3x d^2 \Theta  \,  {\cal X}^{Aa} {\cal X}^{Aa} 
=  \int d^3x \, \left(X_V^{Aa} (F_V^{Aa} - D_6 X_H^{Aa}) - \tf12 \bar\chi_1^{Aa} \chi_1^{Aa} \right)\,. 
\end{eqnarray}
The simplest way to rule out (\ref{Xmasses}), and hence (\ref{CS}) as
well, is to notice that $X_V^{Aa} D_6 X_H^{Aa}$ violates $SU(2)_V
\times SU(2)_H$; eliminating the bulk auxiliary $F_V$ also produces
non-invariant interactions.  A term involving the bulk hyper $\int d^2
\Theta \, {\cal Y}^{Aa} {\cal Y}^{Aa}$ suffers from similar problems.
Finally, one may imagine
\begin{eqnarray}
\label{XY}
S_{XY} &=& \int d^3x d^2 \Theta \, {\cal X}^{Aa} {\cal Y}^{Aa} \\ &=& \int
d^3x \, \tf12 \, \left(X_H^{Aa} (F_V^{Aa} - D_6 X_H^{Aa}) + X_V^{Aa} (F_H^{Aa}
+ D_6 X_V^{Aa}) + \bar\chi_1^{Aa} \chi_2^{Aa} \right)\,.
\nonumber
\end{eqnarray}
Interestingly, almost every bosonic term in $S_{XY}$ is $SU(2)_V
\times SU(2)_H$ invariant; the one exception is a term $(\int d^3x \,
\epsilon_{ABC} f^{abc} X_H^{Aa} X_V^{Bb} X_V^{Cc})$ arising from the
auxiliary fields.  Nonetheless, this term allows us to rule it out.
This completes our list of potential marginal terms with additional couplings.

Since our sought-after field theory must be conformal, we must not
have any massive parameters in the action.  Moreover, for the quantum
theory to maintain conformal symmetry, it is necessary that couplings
of dimension $m$ are not generated by linear divergences.
Consequently, it is useful to demonstrate that mass terms are ruled
out by ${\cal N}=4$ supersymmetry and $SU(2)_V \times SU(2)_H$.  One
might imagine the ${\cal N}=1$ supersymmetric couplings
\begin{eqnarray}
\label{Qmass}
S_{m} &=& \int d^3x d^2 \Theta \, (m \delta_{ij} + m^A \sigma^A_{ij})  \,
\bar{Q}^i Q^j\\ &=& \int d^3x \, (m \delta_{ij} + m^A \sigma^A_{ij} ) \,
(\bar{f}^i q^j + \bar{q}^i f^j - \bar\Psi^i \Psi^j )\,.
\nonumber
\end{eqnarray}
Although the triplet mass term is ${\cal N}=4$ supersymmetric, neither
term is $SU(2)_V \times SU(2)_H$ invariant, since elimination of the
$f^i$ leads not only to $\bar{q}^i q^j$ mass terms, but also to
cross-terms $\sigma^A_{ij} \bar{q}^i X_V^{Aa} T^a q^j$ and $(\sigma^A
\sigma^B)_{ij} \bar{q}^i X_V^{Aa} T^a q^j$, which violate the
R-symmetry.  

Meanwhile, terms involving ambient fields with a massive coupling
constant are impossible, since on dimensional grounds the superspace
integrand would have to contain a single superfield, which cannot be
gauge invariant for $SU(N)$, in the spirit of a defect
Fayet-Iliopoulos term.  Hence the preserved R-symmetry forbids mass
parameters of any kind.

Although we have not imposed them, we find that scale invariance and
parity (\ref{Parity}) are both symmetries of our final classical
action.  (The other discrete symmetry, $P_6$ (\ref{Parity6}), is also
a symmetry, but we have in effect imposed it by demanding that the
defect matter couple only to the ambient vector multiplet.)  Furthermore,
it is also straightforward to show that the action is invariant under
inversion, and hence is fully $SO(3,2)$ symmetric.  Almost all the
rejected couplings would have violated the parity symmetry $P$; this is
not surprising, since 3D fermion mass bilinears are known to violate
parity, and (\ref{SQuartic}) contains an analogous $\bar{q} q
\bar{\Psi} \Psi$.  The exception is (\ref{XY}), which respects parity;
the term $(\int d^3x \, \epsilon_{ABC} f^{abc} X_H^{Aa} X_V^{Bb}
X_V^{Cc})$ is unusual in that it is parity-invariant but $SU(2)_V
\times SU(2)_H$ non-invariant.

We have concluded that our theory is an 3D ${\cal N}=4$
supersymmetric, $SU(2)_V \times SU(2)_H$-invariant coupling of bulk
${\cal N}=4$ Super-Yang Mills to the defect hypermultiplet, also
respecting the $SU(N)$ gauge symmetry and the global $U(1)_B$, and
additionally we were unable to find any further generalizations of the
theory that preserve these symmetries.  Consequently, we conclude that
the action we have obtained defines the correct candidate for a novel
defect superconformal field theory dual to the $AdS_5$/D5-brane
system.  We discuss conformal invariance in the quantum theory in
section~\ref{PerturbSec}.

\setcounter{equation}{0}
\section{Operator matching}
\label{OperatorSec}

The spectrum of modes resulting from the KK reduction of the D5-brane
fields in section~\ref{GravitySec} must be matched with
gauge-invariant operators in the field theory. In this section, we
discuss the construction of this dictionary.  We identify conclusively
the operators dual to the lowest floor of the tower of KK modes.  We
also discuss the primary operators at higher Kaluza-Klein levels.  At
the end, we make a few remarks about the effect of the defect on the
closed-string mode identification.
\begin{center}
\begin{tabular}{c|c|c|c|c|c}
Mode & $m^2$ & $\Delta$ & $SU(2)_H$ & $SU(2)_V$ & Operator in
lowest multiplet\\ \hline $b_\mu$ & $l(l+1)$ & $l+2$ & $l \geq 0$ & $0$ &
$i \bar{q}^m \overleftrightarrow{D^k} \, q^m + \bar\Psi^i \rho^k
\Psi^i$\\ $\psi$ & $(l+2)(l-1)$ & $l+2$ $(1-l)$ & $l \geq 0$ & $1$ &
$\bar\Psi_i \sigma_{ij}^A \Psi_j + 2  \bar{q}^m X_V^{Aa} T^a
q^m$ \\ $(b+z)^{(-)}$ & $l(l-3)$ & $l$ $(3-l)$ & $l \geq 1$ & $0$ &
$\bar{q}^m \sigma_{mn}^I q^n$ \\ $(b+z)^{(+)}$ & $(l+4)(l+1)$ & $l+4$
& $l \geq 0$ & $0$ & ---
\end{tabular}
\end{center}
Above we summarize the results from section~\ref{GravitySec}, 
where in the second and third lines we have noted the possibility of
$\Delta_-$ for small values of $l$.  We have also indicated the three
dual operators that appear in the lowest (massless) multiplet, which
we identify below; the $(b+z)^{(+)}$ tower does not contribute to this
multiplet.

The fields of the dual field theory, their quantum numbers
and their conformal dimensions in the free theory are tabulated below:
\begin{center}
\begin{tabular}{c|c|c|c|c|c}
Mode &  Spin & $SU(2)_H$ & $SU(2)_V$ & $SU(N)$ & $\Delta$ \\ \hline
$A_k$ & $1$ &  $0$ & $0$ & adj & $1$ \\
$X^A_V$ & $0$ &  $0$ & $1$ & adj & $1$ \\
$A_6$ & $0$ &  $0$ & $0$ & adj & $1$ \\
$X^I_H$ & $0$ &  $1$ & $0$ & adj & $1$ \\ 
$\lambda_{im}$ & ${1 \over 2}$ & ${1 \over 2}$ & ${1 \over 2}$ & adj &
${3 \over 2}$ \\
$q^m$ & $0$ & ${1 \over 2}$ & $0$ & {\bf N} & ${1 \over 2}$ \\
$\Psi^i$ & ${1 \over 2}$ & $0$ & ${1 \over 2}$ & {\bf N} & $1$ 
\end{tabular}
\end{center}
The $SU(2)$ quantum numbers are written in a spin notation.  From
these fields, we can construct gauge-invariant operators.  Since the
operators dual to D5-brane modes are confined to the defect, each must
include at least one pair of $q^i$ or $\Psi_a$ fields, but may contain
ambient fields as well.

Certainly it need not be true that every possible operator will have a
dual among the KK SUGRA excitations, as some will instead correspond
to stringy modes, a scenario familiar in $AdS$/CFT.  However, we do
expect to be able to find a dual operator for every D5-brane mode,
because the corresponding multiplets are short, and consequently we
expect the conformal dimensions of the elements to be protected and
not to vary with the 't Hooft coupling $\lambda$. 

In principle, the dual operators are determined by obtaining the full
action for the D3/D5 system before the near-horizon limit is taken.
Terms linear in D5-brane modes then give the composite operator,
composed of both D3 and $3/5$ fields, dual to the D5-brane mode.  We
can deduce the identities of the dual operators in the ground floor by
exploiting supersymmetry alone.  T-duality in the D3/D5 system
provides a check on these results, and identifies the structure of the
higher multiplets.

Consider first the bottom of the $(b+z)^{(-)}$ tower, $l=1$.  This
mode lies in the mass region where either $\Delta_+$ or $\Delta_-$ is
possible.  Since the theory is superconformal and we have the usual
relation between the conformal dimension and the R-symmetry, we
expect $\Delta$ ascend linearly in $l$, and hence we identify the correct
choice as $\Delta_- = 1$.  The operator must be a spacetime scalar in
the $({\bf 3}, {\bf 1})$ of $SU(2)_H \times SU(2)_V$, and there is a
unique candidate:
\begin{eqnarray}
\label{ChiralPrimary}
{\cal C}^I \equiv \bar{q}^m \sigma_{mn}^I q^n \,.
\end{eqnarray}
All the other operators dual to D5-brane modes have larger conformal
dimension, and hence we identify ${\cal C}^I$ as the lowest chiral
primary.  The remainder of the lowest multiplet can be obtained by
acting on ${\cal C}^I$ with ${\cal N}=4$ supersymmetry
transformations.  We can easily do so by beginning with the component
${\cal N}=1$ supersymmetry transformations implicit in
(\ref{SuperTransform}) and promoting the supersymmetry parameter to a
$2 \times 2$ matrix of Majorana spinors $\eta_{im}$, which transforms
like the gaugino $\lambda_{im}$.  We find the other operators in the
same multiplet as ${\cal C}^I$ to be
\begin{eqnarray}
\label{FermiMode}
{\cal F}^{im} &\equiv& {\Psi^*}^i q^m + \bar{q}^m \Psi^i \,, \\
\label{MiddleMode}
{\cal E}^A &\equiv& \bar\Psi_i \sigma_{ij}^A \Psi_j + 2 \,
\bar{q}^m X_V^{Aa} T^a q^m \,, \\
J^k_B &\equiv& i \bar{q}^m D^k q^m - i (D^k q^m)^\dagger q^m + \bar\Psi^i
\rho^k \Psi^i \,,
\label{U1Current}
\end{eqnarray}
where to obtain (\ref{MiddleMode}) we used the explicit form of $f$
(\ref{fSolve}).  We can readily match the bosonic operators ${\cal
E}^A$, $J_B^k$ to D5-brane modes.  ${\cal E}^A$ is an
$SU(2)_V$-triplet and a spacetime scalar with $\Delta = 2$, and hence
matches the $l=0$ mode of $\psi$, assuming we choose $\Delta_+$.
Furthermore, $J^k_B$ is precisely the current of the global symmetry
$U(1)_B$, with $\Delta=2$ and vanishing $SU(2)_V \times SU(2)_H$
quantum numbers, and correspondingly is dual to the lowest mode of
$b_\mu$.

This operator map implies the existence of terms in the action of the
full D3/D5 system, localized on the intersection and coupling the
D5-brane fluctuations to the fields making up the dual operators.  For
example, the identification of (\ref{U1Current}) as the dual of the
D5-brane gauge field implies a coupling
\begin{eqnarray}
\label{D3D5Coupling}
S_{D3/D5} \supset \int d^3x \, B_k \, J^k_B \,,
\end{eqnarray}
which is precisely what we expect given that in the full brane system,
the defect fields are in the fundamental of the D5-brane gauge group
as well.  The supersymmetric partners of (\ref{D3D5Coupling}) must
reproduce the rest of the ground floor operator map.  For us, by far
the easiest way to confirm this is to T-dualize our defect action
(\ref{Skin}), (\ref{SX}) in the 4 and 5 directions; this transforms
the D3-branes into D5-branes and vice versa, and hence generates from
the coupling of D3 fields to the defect the analogous D5-brane
couplings to the defect.  We find that the terms in the dSCFT action
T-dualize to terms that confirm the identification of the operators
(\ref{ChiralPrimary}), (\ref{MiddleMode}), and (\ref{U1Current}).
This agreement is strong evidence that the field theory action we have
developed is the correct candidate for a dual description of the
gravity background.

Let us now consider the higher-$l$ modes.  In analogy with
the usual $AdS$/CFT case, we expect the chiral primary for each value
of $l$ to be obtained from ${\cal C}^I$ by inserting $l$ copies of an
operator ${\cal O}^J$ with $\Delta = 1$ and $SU(2)_H$ spin-1, and
taking the symmetric traceless part:
\begin{eqnarray}
\label{HigherChiralPrimary}
C_l^{I_0 \ldots I_l} = C^{(I_0} {\cal O}^{I_1} \ldots {\cal O}^{I_l)} \,. 
\end{eqnarray}
In principle the quantum numbers permit two candidates for ${\cal
O}^I$:
\begin{eqnarray}
\bar{q}^m \sigma_{mn}^I q^n \,, \quad \quad X_H^I \,.
\end{eqnarray}
From the point of view of the intersecting brane system, $X_H^I$ is
the natural choice to generate higher moments of D5-brane fields.  On
the other hand, one might worry that $X_H^I$ is an unnatural candidate
for an operator that generates chiral primaries, since it is a member
of the inert bulk hypermultiplet that does not even couple to the
defect fields.  One can once again turn to T-duality in the full brane
system to argue that $X_H^I$ is the right choice.

To do so, one must notice an additional constraint on possible terms
localized on the intersection in the D3/D5 system.  T-duality along
the 4 and 5 directions carries the system into itself, so the total
set of these terms must be invariant up to a relabeling of
coordinates.  However, this operation interchanges indices of D3 or
D5-brane modes polarized on $I=345$ with those polarized in the
$6$-direction.  Consequently, a generic term that is
$SU(2)_H$-invariant before T-duality might not be afterwards; such
terms cannot be present in the brane action.  In order to reconcile
T-duality with $SU(2)_H$, one must require that an even number of D3
or D5 indices in either the $345$ or $6$ directions appear.  This
constraint turns out to be equivalent to the requirement that the set
of D3-brane and D5-brane ambient hypermultiplet fields only appear in
pairs.

Up until now we have not discussed the $l=0$ mode at the bottom of the
$(b+z)^{+}$ tower, which is simply the constant mode of $Z^6$ with no
$b_\alpha$ contribution. This mode appears in the second floor short
multiplet, and is dual to an R-singlet operator ${\cal O}_{Z^6}$ with
$\Delta=4$.  Hence there must exist a coupling in the D3/D5 brane system
\begin{eqnarray}
S_{D3/D5} \supset \int d^3x \, \alpha' \, Z^6 \, {\cal O}_{Z^6} \,.
\end{eqnarray}
Unlike the D5-brane modes appearing in the ground floor
operator map, $Z^6$ is in the D5-brane ambient hypermultiplet, not the
vector.  T-duality hence demands that at least one D3-brane ambient
hyper field appear in ${\cal O}_{Z^6}$.  Now, ${\cal O}_{Z^6}$ must be a
four-supercharge descendant of the second-floor chiral primary
$C_1^{IJ}$; however, one may show that no such descendant of $\bar{q}
\sigma^{(I} q \bar{q} \sigma^{J)} q$ contains a D3 hyper field.  On
the other hand, $X_H^I$ is itself in the D3-brane ambient hyper, and
$\bar{q} \sigma^{(A} X_H^{B)} q$ indeed does have descendants
containing such a field.

Hence, we identify $\bar{q} \sigma^{(A} X_H^{B)} q$ as the consistent
choice for the second-floor chiral primary, and $X_H^J$ as the
operator ${\cal O}^J$ that generates all higher chiral primaries $C_l$
corresponding to the $(b+z)^-$ tower as (\ref{HigherChiralPrimary}).
The operators dual to the remaining D5-brane modes, including ${\cal
O}_{Z^6}$ and its higher moments, can be obtained from the $C_l$ by
supersymmetry transformations.  This determines in principle the complete
D5-brane mode/defect operator dictionary.

Before turning to perturbative calculations, let us mention the effect
of the defect on the closed-string mode/operator dictionary.  The
leading-order identification of bulk closed string fields to operators
varying over the ambient 4D space will remain unchanged, but
corrections can arise localized on the defect.  One obvious example of
this is the energy-momentum tensor, dual to the transverse traceless
graviton, which has the form
\begin{eqnarray}
T_{\mu \nu} = T^{{\cal N}=4}_{\mu \nu} + \delta(x) \, T^{3d}_{kl} \,
\delta^k_{\; \mu} \, \delta^l_{\; \nu} \,.
\end{eqnarray}
Note that tracelessness of the full stress tensor, associated with
conformal invariance, refers to a trace over all 4 indices, not just
3, despite the fact that the conformal group is just $SO(3,2)$.  This
reflects the fact that the realization of scale transformations is
four-dimensional, reducing to a 3D scale transformation only on the
defect.  

The dilaton, which is the supersymmetric partner of the graviton,
should be dual to the total field theory Lagrangian, including defect
terms.  Similarly, other operators in the same reduced supersymmetry
multiplet may have a $\delta(x)$ piece.  Obtaining the contributions of
such defect pieces to correlation functions via gravity calculations
is an open problem.  Some bulk modes, such as the scalars dual to
${\rm Tr}\, X^2$, lie in different multiplets; whether they also
receive a localized part at leading order would be interesting to
determine.

\setcounter{equation}{0}
\section{Perturbative Field Theory}
\label{PerturbSec}

There is by now a vast literature discussing the interactions of
matter localized on a boundary with higher-dimensional fields, chiefly
inspired by the Ho\v{r}ava-Witten scenario \cite{HoW} and involving a
five-dimensional bulk caught between two ``end-of-the-world''
3-branes.  Perturbative analysis for such theories in a spirit similar
to this paper can be found in \cite{Peskin}, \cite{Georgi}, \cite{GW}.

The dSCFT dual to the Karch-Randall system is novel in a number of
ways.  First, space does not terminate at the defect but instead
continues through it, and consequently no boundary conditions are
imposed on the ambient fields.  Second, since the total dimension is
four, the gauge theory is renormalizable and hence well-defined in the
ultraviolet.  Finally, despite the presence of the defect, the theory
is postulated to be exactly superconformal.

In this section, we discuss the results of a preliminary study of
the perturbative properties of such theories.  The first task of such
a study should be to investigate whether the classical $SO(3,2)$
conformal symmetry is maintained in perturbation theory, and an
approach to this question is presented in the next subsection. This is
followed by a discussion of weak coupling properties of correlation
functions of composite operators which illuminate issues which arose in
our discussion of the putative gravity dual.

\subsection{Quantum Conformal Invariance?}
\label{ConformalSec}

The elementary yet essential aspect of our defect theories is that
certain fields of the ambient ${\cal N}=4$ SYM theory are ``pinned''
to the defect at $x=0$ and couple as 3-dimensional fields with scale
dimension enhanced by one unit. Thus for a scalar boson $X(x,\vec{y})$
or restricted spinor $\lambda_1(x,\vec{y})$ we have the pinned
propagators (in Euclidean signature)
\begin{eqnarray}
\langle X(0,\vec{y})\, X(0,\vec{y'})\rangle &=& { 1 \over 4\pi^2(\vec{y}-\vec{y}')^2}
                                         = FT_3 \left( {1 \over 2|\vec{k}|} \right) \,, \\
\langle \lambda_1(0,\vec{y}) \, \bar\lambda_1(0,\vec{y'})\rangle
  &=& - {\rho^k(y-y')_k \over 2\pi^2 (\vec{y}-\vec{y}')^4}
  = FT_3 \left({i \rho^k k_k \over 2 |\vec{k}|} \right) \,,
\end{eqnarray}
whereas propagators of defect fields are
\begin{eqnarray}
\langle q(\vec{y}) \, \bar{q}(\vec{y}') \rangle &=& {1 \over
4\pi|\vec{y}-\vec{y}'|} = FT_3 \left( {1 \over \vec{k}^2} \right) \, \\
\langle \Psi(\vec{y}) \,
\bar\Psi(\vec{y}')\rangle &=& -{ \rho^k(y-y')_k \over 4\pi
|\vec{y}-\vec{y}'|^3} = FT_3 \left( {i \rho^k k_k \over \vec{k}^2} \right) \,.
\end{eqnarray}
Of course it is the 3-dimensional Fourier transform, $FT_3 (f(\vec{k})) =
\int d^3k \, e^{i\vec{k}\cdot\vec{y}} \, f(\vec{k})/(2\pi)^3,$ which is
relevant for correlation functions with all external operators pinned
at the defect. We thus find that pinned propagators are more singular
at short distance or high momentum than is standard in 3 dimensions.
It is in this way that the defect theory, which would have been
super-renormalizable if purely in 3 dimensions, becomes critically
renormalizable with dimensionless couplings.

We now outline an argument based on power counting and symmetries that
conformal symmetry is maintained in perturbation theory. We will argue
that, after cancellation among the graphs of fixed loop order
contributing to a given 1PI amplitude, the only new divergences are
those of wave function renormalization of the defect fields $q^i,
\Psi^i,f^i$. Wave function renormalization induces anomalous
dimensions of the elementary fields, which are generically gauge
dependent and non-observable, and thus have no effect on conformal
symmetry.\footnote{Inspection of the unique 1-loop graph for the $f^i$
self-energy reveals immediately that it is logarithmically
divergent. The same is true for the $F^A$ self-energy in conventional
component $D=4$ ${\cal N}=4$ SYM theory in Wess-Zumino gauge.} Our
discussion assumes that the supersymmetry and other symmetries
(e.g. parity and $SU(2)_H \times SU(2)_V$) are maintained in
perturbation theory.

Amputated $n$-point functions of ambient fields generically contain
two types of contributions (for each bulk line) --- a pinned
contribution in which the ``first interaction'' of the external field
is on the defect and an unpinned contribution in which the first
interaction is in the ambient ${\bf R}^4$. Our discussion deals first
with the pinned contributions, which carry an explicit $\delta(x)$
factor.\footnote{For an external $X_H$ line the factor is
$\delta'(x)$.} Divergences of these contributions would require local
counterterms $\delta L= \int d^3y \, {\cal O}_3$ on the defect.
Further the pinned pieces are the only contributions if the ambient 4D
theory is free, e.~g.\ for ${\cal N}=4$ SYM with gauge group $U(1)$.

Let us write a power-counting formula for a generic amputated
$n$-point function with $n_q$, $n_{\Psi}$, $n_f$ external defect
fields and $n_A$, $n_\lambda$, $n_{X_V}$, $n_{X_H}$, $n_\chi$, $n_F$
pinned ambient fields. With modest work, one can see that the
superficial degree of divergence is
\begin{eqnarray}
\delta =2n_A + \tf32 \, n_{\lambda} +2n_{X_V} + n_{X_H} +\tf32 \, n_{\chi} +n_F +\tf52 \, n_q +2 n_{\Psi} + \tf32 \, n_f - 3n +3   \,.
\end{eqnarray}
There is a long list of divergent component amplitudes, of which we
discuss a few in order to convey the essential part of our argument.

Beginning with two-point functions, we see that the $\Psi$ self-energy
is linearly divergent, threatening an infinite mass counterterm
$\bar\Psi^i\Psi^i$.  However, we have pointed out in
Sec.~\ref{ActionSec} that this term is parity violating and, due to
${\cal N}=1$ SUSY, must be accompanied by other terms which are
non-invariant under $SU(2)_H \times SU(2)_V$. Thus the potential
divergence must cancel, and SUSY then implies that the only divergence
of 2-point functions of defect fields is logarithmically divergent
wavefunction renormalization.

We may also consider the effect of the defect on the self-energy of
bulk fields.  The vacuum polarization of the gauge field determines
the renormalization of the coupling $g$.  The contribution to this
quantity with both external fields pinned has linear superficial
degree of divergence, but this is decreased due to gauge
invariance. Decreasing by a single power of external $p_i$ suggests a
log-divergent Chern-Simons counterterm, but this is again prohibited
by parity symmetry, as is the companion pinned mass term
$\bar\lambda_1 \lambda_1$. Thus both the pinned vacuum polarization
and $\lambda_1$ self-energy are UV finite; $SU(2)_H \times SU(2)_V$
symmetry requires the $\chi_1$ self-energy to be finite as well. The
$X_V$ self-energy is linearly divergent, but there is no
Lorentz-invariant $X_V \, \partial_i X_V$ counter term.  The term $X_V
\, \partial_6 X_V$ is Lorentz-invariant and parity-invariant, but it
violates $P_6$, and once again SUSY requires it to appear with other
terms (\ref{XY}) that violate $SU(2)_V \times SU(2)_H$.

Moving on to 3-point functions, we see that the amputated correlator
$\langle A_i \Psi \bar\Psi \rangle$ with the gauge field pinned is log
divergent by power counting.  Although confined to the defect,
the hyperino $\Psi^i$ is a canonically coupled field in the
fundamental of the gauge group, and the usual gauge Ward identity
implies that this divergence is canceled by wavefunction
renormalization. The gauge coupling $g$ can only be renormalized in
the vacuum polarization, for which the defect contribution was argued
to be finite above.  This argument applies not just to our theory, but
to a general coupling of a 4D gauge theory to 3D matter.  In our case,
however, ${\cal N}=4$ SUSY and $SU(2)_H \times SU(2)_V$ invariance
then imply that there are no infinite counterterms for any of the
cubic couplings in $S_{kin}$ (\ref{Skin}), or $S_X$ (\ref{SX}).

The quartic couplings of the scalar potential in $S_{pot}$
(\ref{Spot}) are generated from three-point couplings by eliminating
auxiliary fields, and hence these are also fixed by SUSY and cannot be
renormalized. It has also been shown in Sec~\ref{ActionSec} that other
potentially log divergent $n$-point functions with $n \ge 4$, such as
$\langle \bar{q}^i q^i \bar\Psi^j \Psi^j \rangle$ and $\langle
\bar{q}^i q^i \bar{q}^j q^j \bar{q}^k q^k \rangle$, cannot induce new
couplings because they violate the symmetries.

These remarks add up to a strongly suggestive argument that at least
the diagrams involving defect and pinned ambient fields respect
conformal invariance.  This is sufficient to guarantee conformality
for the $U(1)$ version of our theory, where the gauge charge appears
only in defect interactions.  The gravity dual requires $SU(N)$ gauge
group for the field theory. This necessarily involves non-pinned
contributions to correlators involving both ambient and defect
fields. They are more divergent at short distance, and lack
conventional translation symmetry.  Further study is needed to handle
them.  Thus, although we are optimistic, it is too early to declare
victory on the question of conformal symmetry of the $SU(N)$ theory.

Gauge anomalies can be shown to be absent.  Our theory is still
four-dimensionally gauge invariant, as it must be to make sense of the
4D gauge field, and bulk fields in principle contribute to a 4D gauge
anomaly, which for ${\cal N}=4$ SYM cancels.  Defect fields, however,
participate only in a restricted three-dimensional gauge invariance.
There are no ordinary gauge anomalies in three dimensions.
Three-dimensional theories can possess a parity anomaly that induces a
3D Chern-Simons term \cite{ParityAnomaly}, but this arises only when
there is an odd number of charged Majorana spinors, so our theory is
safe.

One novel feature of the defect theory is the $\delta(0)$ from the
$\bar{q}\, F_R\, q$ vertex in the action. We now give a general argument that
this is a harmless artifact. We start at the level of elementary
auxiliary fields. The propagator of $F_R$ in (Euclidean) momentum space
is $-1$ and that of $X_H$ is $1/(k^2 +\vec{k}^2).$ Thus for
$\partial_6 X_H(x,\vec{y})$, it is $k^2/(k^2 +\vec{k}^2).$ In exchanges
between $\bar{q}q$ pairs one then has the effective propagator
\begin{eqnarray}
\int \frac{dk}{2\pi} \left[1 -\frac{k^2}{k^2 +\vec{k}^2} \right]=
\int \frac{dk}{2\pi}\frac{\vec{k}^2}{k^2 +\vec{k}^2} = |\vec{k}| \,.
\end{eqnarray}
In position space this amounts to the propagator
$1/(2\pi^2)(\vec{y}-\vec{y'})^4$ between $\bar{q}q$ vertices at $(0,\vec{y})$
and $(0,\vec{y'})$. After elimination of $F_R$ one can see that the correct
perturbation expansion is obtained if one neglects the  $\delta(0)$ term in
(\ref{Spot}) and uses the effective propagators above for exchange of
$\partial_6 X_H$ between $\bar{q}q$ pairs. Needless to say this is true for
both tree level exchange, as discussed in \cite{Peskin}, and when the
exchange is included with another amplitude. This suggests that it may be
useful to use a  ${\cal N}=1$ supergraph formalism in which the cancellation
above is automatic.

\subsection{Composite Correlators at Weak Coupling}
\label{CompositeSec}

One clear prediction of the extended $AdS$/CFT correspondence we are
investigating is that a large set of defect operators in the dual
field theory have integer scale dimension. Assuming the conjectured
conformal symmetry is valid, the reason is that these operators span
a short representation of the superalgebra $OSp(4|4)$. It is then
valid to map fields on $AdS_4$ to composite operators on the defect
according to the free field scale dimension of the latter, and this
was done in Sec. 5. Although one would not expect symmetry relations to
fail, it would be desirable to use weak coupling calculations is
to test that radiative corrections to these $\Delta$'s vanish. Although
the $AdS$/CFT duality predicts that most correlators are renormalized,
it is not excluded that 2-point functions of defect operators,
$\langle {\cal O}_3 \, {\cal O}_3 \rangle$, have no radiative corrections.
However, to test these features requires more precise calculations than
time has so far allowed us.

It is nevertheless possible to use weak coupling analysis to
illuminate some aspects of the operator map and we now discuss one
application.  Kaluza-Klein analysis led us to a unique operator of
dimension $\Delta =1$ in the open string/defect operator dictionary,
namely the $SU(2)_H$ triplet ${\cal C}^A \equiv \bar{q} \sigma^A q$ of
(\ref {ChiralPrimary}).  The singlet $\bar{q}q$ is not in the operator
map. Generically one would expect it to have anomalous dimension, and
we will show that this does happen to order $g^2N$. 

\begin{figure}
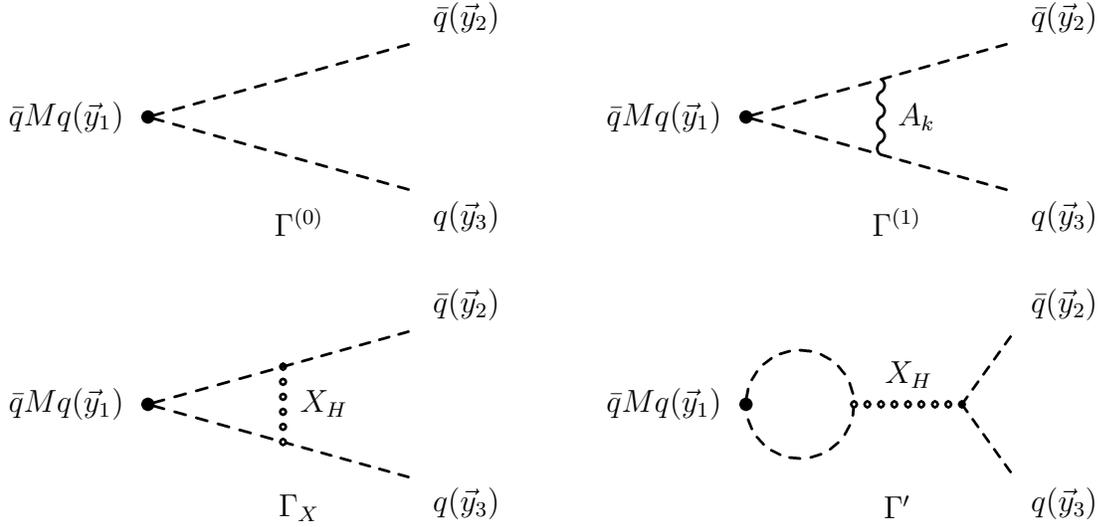

\begin{center}
\begin{fmfchar*}(40,20)
  \fmfleft{i} \fmflabel{$\bar{q}Mq (\vec{y}_1)\;$}{i}
  \fmfrightn{o}{2}
  \fmfbottom{b}
  \fmf{dashes}{i,o1}
  \fmf{dashes}{i,o2}
  \fmflabel{$\bar{q}(\vec{y}_2)$}{o2} \fmflabel{$q(\vec{y}_3)$}{o1}
  \fmfdot{i}
  \fmflabel{$\Gamma^{(0)}$}{b}
\end{fmfchar*}
\hskip1.5in
\begin{fmfchar*}(40,20)
  \fmfleft{i} \fmflabel{$\bar{q}Mq (\vec{y}_1)\;$}{i}
  \fmfrightn{o}{2}
  \fmfbottom{b}
  \fmf{dashes}{i,j1,o1}
  \fmf{dashes}{i,j2,o2}
  \fmffreeze
  \fmf{photon,label=$A_k$}{j1,j2}
  \fmflabel{$\bar{q}(\vec{y}_2)$}{o2} \fmflabel{$q(\vec{y}_3)$}{o1}
  \fmfdot{i}
  \fmflabel{$\Gamma^{(1)}$}{b}
\end{fmfchar*}\break
\vskip.5in
\begin{fmfchar*}(40,20)
  \fmfleft{i} \fmflabel{$\bar{q}Mq (\vec{y}_1)\;$}{i}
  \fmfrightn{o}{2}
  \fmfbottom{b}
  \fmf{dashes}{i,j1,o1}
  \fmf{dashes}{i,j2,o2}
  \fmffreeze
  \fmf{dbl_dots,label=$X_H$}{j1,j2}
  \fmflabel{$\bar{q}(\vec{y}_2)$}{o2} \fmflabel{$q(\vec{y}_3)$}{o1}
  \fmfdot{i}
  \fmflabel{$\Gamma_X$}{b}
\end{fmfchar*}
\hskip1.5in
\begin{fmfchar*}(40,20)
  \fmfleft{i} \fmflabel{$\bar{q}Mq (\vec{y}_1)\;$}{i}
  \fmfrightn{o}{2}
  \fmfbottom{b}
  \fmf{dashes,left,tension=.5}{i,j,i}
  \fmf{dbl_dots,label=$X_H$,lab.sid=left}{j,k}
  \fmf{dashes}{k,o1}
  \fmf{dashes}{k,o2}
  \fmfdot{i}
  \fmflabel{$\Gamma'$}{b}
  \fmflabel{$\bar{q}(\vec{y}_2)$}{o2} \fmflabel{$q(\vec{y}_3)$}{o1}
\end{fmfchar*}
\vskip.3in
\end{center}
\caption{Feynman diagrams through one-loop order for the correlators
$\langle \bar{q} M q(\vec{y}_1) \, \bar{q}(\vec{y}_2) \, q(\vec{y}_3)
\rangle$ where $M$ is either $\sigma^A$ or $1$.}
\label{Feynman}
\end{figure}

The operator $\bar{q} \sigma^A q$ is the primary of the multiplet
containing the conserved current $J_B^k$ of (\ref{U1Current}), so it
is fair to assume that its scale dimension is exactly $\Delta
=1$. Given this assumption it is not difficult to compare graphs for
the 3-point functions $\langle \bar{q} \sigma^A q(\vec{y}_1) \,
\bar{q}(\vec{y}_2) \, q(\vec{y}_3) \rangle$ and $\langle
\bar{q}q(\vec{y}_1) \, \bar{q}(\vec{y}_2) \, q(\vec{y}_3) \rangle$
through 1-loop order and show that $\bar{q}q$ acquires anomalous
dimension. We work implicitly in the framework of differential
regularization \cite{FJL} in which no counterterms are needed and
renormalization data is inferred directly from the Callan-Symanzik
equations
\begin{eqnarray}
\left [M{d \over dM} +\beta(g) {d \over dg} -2 \gamma_q -
\gamma_{{\cal O}} \right] \langle {\cal O}(\vec{y}_1) \,
\bar{q}(\vec{y}_2) \, q(\vec{y}_3) \rangle =0 \,.
\end{eqnarray}
The two 2-point functions can be expressed as follows:
\begin{eqnarray}
\langle \bar{q}q(\vec{y}_1) \, \bar{q}_i(\vec{y}_2)
q_j (\vec{y}_3)\rangle &=& \delta_{ij}[\Gamma +3\Gamma_X] \,, \\
\langle \bar{q}\sigma^A q(\vec{y}_1) \,  
\bar{q}_i(\vec{y}_2)  q_j(\vec{y}_3) \rangle &=& \sigma^A_{ij}
[\Gamma + \Gamma' -\Gamma_X] \,.
\end{eqnarray}
The Feynman diagrams which contribute to $\Gamma = \Gamma^{(0)} +
\Gamma^{(1)}, \Gamma', \Gamma_X$ are given in Fig.~\ref{Feynman}. The
$SU(2)_H$ algebra for these diagrams has been done and incorporated in
the equations above, while color is suppressed. The analysis succeeds
because the $X_H$ exchange diagram $\Gamma_X$ has different weights in
the two amplitudes.

We have argued in Sec~\ref{ConformalSec} that $\beta(g)=0$, but, even
if not, the lowest order contribution is $\beta \sim g^3$ which cannot
affect the present argument. Writing $\Gamma^{(0)},\Gamma^{(1)}$ to
distinguish tree and 1-loop contributions to $\Gamma$, the
perturbative CS equations can be written as:
\begin{eqnarray}
M{ d \over dM}(\Gamma^{(1)} + 3 \Gamma_X) &=&
(2\gamma_q+\gamma_{\bar{q}q})\Gamma^{(0)} \,, \\ M{ d \over
dM}(\Gamma^{(1)}+\Gamma'-\Gamma_X) &=& (2\gamma_q+\gamma_{\bar{q}\sigma
q})\ \Gamma^{(0)}\ \,.
\end{eqnarray}
The graph $\Gamma'$ is UV finite (it turns out to be a numerical multiple
of $g^2\Gamma^{(0)}$), and its scale derivative thus vanishes. However, both
$\Gamma^{(1)}$ and $\Gamma_X$ are log divergent. By subtraction,
the two equations then give
\begin{eqnarray}
4 M{d \over dM} \Gamma_X = (\gamma_{\bar{q}q} -\gamma_{\bar{q}\sigma
q})\Gamma^{(0)} \,.
\end{eqnarray}
If $\gamma_{\bar{q}\sigma q}$ vanishes, as we assume,
$\gamma_{\bar{q}q}$ is non-vanishing. Thus $\bar{q}q$ has radiatively
corrected scale dimension $\Delta_{\bar{q} q} = 1 + \gamma_{\bar{q}q}$.

Another application of perturbative analysis to the operator map is
studying the two candidate operators discussed in
Sec.~\ref{OperatorSec} whose multiple products might appear as field
theory duals of higher D5-brane KK fluctuations on $AdS_4$.  The
chiral primary fields of the KK multiplets are modes of $(b+z)^{(-)}$
with $SU(2)_H$ quantum number $\ell$ (with $\ell \ge 2$) and scale
dimension $\Delta=\ell$. The two families of candidate dual operators
are the isospin $\ell$ components of $(\bar{q} \sigma^A q)^{\ell}$ and
those of $\bar{q} \sigma^A (X_H^B)^{(\ell-1)} q$.  In
Sec.~\ref{OperatorSec} we presented an argument based on the T-duality
invariance of the defect D3/D5 action suggesting that the latter
family is the correct choice.  We will now outline an argument based
on the Callan-Symanzik equation which shows that the former set of
operators has no anomalous dimension to lowest order. The virtue of
this argument, which is similar to that for $\bar{q} q$ above, is that
a precise evaluation of the diagrams is not required. This is not true
for the operator family $\bar{q} \sigma^A (X_H^B)^{(\ell-1)} q$ since
there are more contributing diagrams, so the question of anomalous
dimension for these is not yet settled.

We choose highest weight component of the $\ell =2$ projection of
$(\bar{q} \sigma^A q)^{2}$ and study all tree and 1-loop graphs for
the 5-point function $\langle \bar{q}^1 q^2 \bar{q}^1 q^2(\vec{y}) \,
q^1(\vec{y_1}) \, \bar{q}^2(\vec{y}_2) \, q_1(\vec{y}_3) \,
\bar{q}^2(\vec{y_4})\rangle$. There are 1-loop graphs with gluon and
$\partial_6X_H$ exchange between the $q$-lines at $\vec{y}_1$ and
$\vec{y}_2$. These graphs contribute no anomalous dimension in the CS
equation since they enter in the same way as for the protected
operator $\bar{q} \sigma^A q$. The same is true for exchanges between
lines at $\vec{y}_3$ and $\vec{y}_4$. There are additional UV finite
graphs as in $\Gamma'$ above. There remain 4 graphs with gluon
exchange between $\vec{y}_1$ or $\vec{y_2}$ and $\vec{y}_3$ or
$\vec{y}_4$ and 4 more graphs with exchange of $\partial_6 X_H$. The
amplitudes of the graphs are not the same space-time functions, but
their contribution to the scale derivative is proportional to the same
local tree amplitude in all cases. There are two gluon exchanges
between $qq$ and two between $q\bar{q}$. Coefficients are equal and
opposite and the sum cancels. One can examine the $SU(2)_H$ flavor
algebra and find a similar cancellation among the 4 $\partial_6 X_H$
exchange graphs. In this way we have shown that the 5-point function
satisfies the CS equation with no order $g^2N$ anomalous dimension for
the $\ell=2$ components of the operator $(\bar{q} \sigma^A q)^2$. The
same argument fails for $\ell=0,1$ components because there are
inequivalent color contractions.

It is a matter of simple combinatorics to extend the argument to the
highest weight $\ell=n$ components of $(\bar{q} \sigma^A q)^n$. One first
separates graphs with interactions on $q$-lines which terminate at
a single $\bar{q}\sigma^A q$ factor in the product. These graphs do not
contribute to the anomalous dimension, as above. There remain $2(n-1)$
gluon exchanges between $q\bar{q}$ and  $2(n-1)$  between $qq$. Their
contribution to the scale derivative cancels as above. Finally, there
are $4(n-1)~\partial_6X_H$ exchanges. Within groups of 4 one can study the
flavor algebra and find complete cancellation. 

\begin{figure}
\begin{center}
\begin{fmfgraph*}(45,30)
  \fmfleft{r} \fmfright{q}  \fmflabel{${\rm Tr}\, (X_V)^3$}{r}
  \fmfdot{r}
  \fmfipair{i,u,c,d,tu,td}
  \fmfiequ{i}{.5[sw,nw]}
  \fmfiequ{c}{.5[se,ne]}
  \fmfiequ{u}{.8[se,ne]}
  \fmfiequ{d}{.2[se,ne]}
  \fmfblob{.41w}{q}
    \fmfi{dbl_dots,label=$X_V$}{i{ne} ..  tension 1.8 .. {right}u}
    \fmfi{dbl_dots,label=$X_V$}{i{sw} ..  tension 1 .. {right}d}
    \fmf{dbl_dots,label=$X_V$}{r,q}
    \fmfi{plain}{ne--se}
\end{fmfgraph*}
\end{center}
\caption{The generic contribution to the one point function $\langle
{\rm Tr}\, (X_V)^k \rangle$ with all $k$ $X_V$ lines pinned to
the defect (vertical line), depicted here for $k=3$.}
\label{Feynman2}
\end{figure}
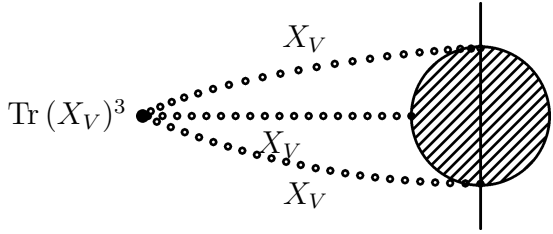

We conclude our survey of perturbative results with a discussion of
the field theory interpretation of poles that appeared in the gravity
calculations of section \ref{GravitySec}.  In the computation of
$\langle {\cal O}_4 \rangle$ from the D5-brane action, we noted a
divergence for $\Delta_4 \le 3$ which comes from the boundary region
of the integration over $AdS_4$.  In the conventional $AdS$/CFT
correspondence similar infinities can be interpreted as UV divergences
in the dual field theory. A parallel interpretation seems plausible
here.  For ${\cal O}_4= {\rm Tr}\, (X_V)^k$ some Feynman diagrams
contain a generic sub-amplitude with $k$ pinned $X_V$ lines (as shown
in Fig.~\ref{Feynman2} for $k=3$). The degree of divergence is
$\delta= 3-k$.  Thus the diagram has a subdivergence (as all
interaction points $\vec{y_I}$ on the loop come together) for $k \le
3$ in perfect correspondence with the gravity result. Of course the
divergence on the gravity side is present for generic $AdS_4$ action,
but cancels due to symmetry in our specific case. In field theory as
well, the divergence predicted by generic power counting also violates
symmetry and cancels.  For the case $k=2$ the field theory amplitude
is linear divergent, but the gravity result is finite. However, in low
order examples, the divergence cancels due to symmetric integration
leaving a finite remainder. One may also apply similar power counting
to field theory amplitudes for $\langle {\cal O}_4 \, {\cal O}_3
\rangle$ and find that a subdivergence is formally predicted when
$\Delta_3-\Delta_4 \ge 0$ in agreement with the calculation in
supergravity.

\section{Open questions}

Many avenues remain for further exploration.  The most pressing issue
is the proof of conformality for $SU(N)$ gauge group.  Assuming that
the theory is conformal, one is naturally led to wonder about the
existence of other dSCFTs.  Simple generalizations include changing
the gauge group, the defect matter representation, or promoting
$U(1)_B$ to a nonabelian symmetry; this last possibility may be
holographically related to a theory with multiple D5-branes.
Completely different dSCFTs in other dimensions likely exist as well,
and may have holographic duals.

A more detailed study of the correlation functions of the field theory
described in the present paper would also be interesting, including a
precise matching with results from the D5-brane action containing KK
reduced bulk modes.  The question of the existence of
non-renormalization theorems for correlators with two defect operators
should be investigated.  There also remains the more general
understanding of how the presence of the defect corrects the closed
string/ambient operator map and the related correlation functions.
Whether the gravity coupling vanishes for ``extremal'' correlators $\langle
{\cal O}_4 \, {\cal O}_3 \rangle$ is a test of our reasoning
concerning the pole structure.

The importance of determining the supergravity solution taking account
of the back-reaction of many D5-branes, as emphasized by \cite{KR1},
remains.  Such a geometry must produce all the physics of the dSCFT
through closed string excitations alone, presumably by means of local
localization.  Finally, it would be fascinating to deform this
correspondence away from the conformal limit, and to study the
holographic duality between the much broader class of defect field
theories that run with scale and more intricate brane geometries.

\medskip \noindent {\em Note added} \quad A demonstration of the conformality
of the full non-Abelian theory has recently appeared in \cite{EGK}.

\section*{Acknowledgments}

We are grateful for conversations with Allan Adams, Costas Bachas,
Alex Buchel, Jan de Boer, Robbert Dijkgraaf, Noah Graham, Ami Hanany,
Petr Ho\v{r}ava, Andreas Karch, Igor Klebanov, Andrei Mikhailov, Joe
Polchinski, Kostas Skenderis, Witek Skiba, Jan Troost and Wati Taylor.
We also thank Sergey Frolov and Massimo Porrati for pointing out minor
errors in the first version.  The work of O.D.\ was supported by the NSF under grant
PHY-99-07949.  The work of D.Z.F.\ was supported by the NSF under
grant PHY-00-96515.  The work of H.O.\ was supported in part by DOE
grant DE-AC03-76SF000098. D.Z.F.\ and H.O.\ would also like to thank
ITP, Santa Barbara for hospitality.

\section*{Appendix A: Spherical Harmonics on $S^5$ and $S^2$}

Bulk fields are expanded in spherical harmonics on the $S^5$.  For
example, for scalar harmonics, we can write the $S^5$ harmonics in
terms of products of standard harmonics $Y^\ell_m(\theta,\varphi)$,
$Y^{\ell'}_{m'}(\chi,\varsigma)$ on each $S^2$ and functions of the fifth
coordinate $\psi$:
\begin{eqnarray}
Y^k_{\ell m \ell' m'}(\psi,\theta,\varphi,\chi,\varsigma) =
Y^\ell_m(\theta,\varphi) Y^{\ell'}_{m'}(\chi,\varsigma) Z^k_{\ell
\ell'}(\psi) \,,
\end{eqnarray}
The $S^5$ Laplacian in the coordinates (\ref{SphereMetric}) is
\begin{eqnarray}
\label{SphereLaplacian}
\square_{S^5} = {1 \over \sin^2 \psi \cos^2 \psi} {\partial \over
\partial \psi} \sin^2 \psi \cos^2 \psi {\partial \over \partial \psi}
+ {1 \over \cos^2 \psi} \square_{\theta, \varphi} + {1 \over \sin^2
\psi} \square_{\chi, \varsigma} \,.
\end{eqnarray}
A scalar spherical harmonic $Y^k$ on $S^q$ transforms in the $k$-fold
symmetrized traceless product of fundamentals of $SO(q+1)$.  It is an
eigenvalue of the Laplacian on $S^q$ with eigenvalue
\begin{eqnarray}
\square_{S^q} Y^k = - k (k+q-1) Y^k \,,
\end{eqnarray} 
and using (\ref{SphereLaplacian}) we can obtain an ordinary
differential equation for $Z^k_{\ell \ell'}(\psi)$,
\begin{eqnarray}
\label{ZDiffEq}
\left( {1 \over \sin^2 \psi \cos^2 \psi} {\partial \over
\partial \psi} \sin^2 \psi \cos^2 \psi {\partial \over \partial \psi}
- {\ell(\ell +1) \over \cos^2 \psi} - { \ell' (\ell' +1)\over \sin^2
\psi} \right) Z^k_{\ell \ell'} (\psi) = - k(k+4) Z^k_{\ell \ell'} (\psi) \,.
\end{eqnarray}
Since there are interactions between closed-string and D5-brane fields
on the D5 worldvolume, we are interested in the behavior of the
spherical harmonics at $\psi = 0$.  To leading order in $\psi$, the
equation for $Z^k_{\ell \ell'}(\psi)$ (\ref{ZDiffEq}) reduces to
\begin{eqnarray}
\label{ZDiffEq2}
\left( {\partial^2 \over \partial \psi^2} + {2 \over \psi} {\partial
\over \partial \psi} - { \ell' (\ell' +1)\over \psi^2} +k(k+4)-
\ell(\ell +1) \right) Z^k_{\ell \ell'} (\psi) = 0 \,.
\end{eqnarray}
We perform a standard Frobenius analysis by expanding $Z^k_{\ell
\ell'}(\psi)$ near $\psi = 0$ as $Z(\psi) = \psi^\alpha
\sum_{n=0}^\infty x_n \psi^n$, where we are always free to take $x_0
\neq 0$ by redefining $\alpha$ if necessary.  The leading order
term in (\ref{ZDiffEq2}) then leads to the requirement
\begin{eqnarray}
\label{HarmScaling}
\alpha = \ell' \quad \quad {\rm or} \quad \quad \alpha = - \ell' -1 \,.
\end{eqnarray}
Requiring the regularity of the spherical harmonics over the complete
$S^5$ selects the former.  We are then led to the conclusion that
$Z^k_{\ell \ell'}(\psi = 0) = 0 $, and by extension $Y^k_{\ell m \ell'
m'}(\psi = 0) = 0$, unless $\ell' = 0$. We conclude that only the
closed-string modes invariant under $SU(2)_V$ couple directly to the
D5-brane fields.  

Another conclusion we can draw is that given $\ell$ and $\ell'$, there
no more than one harmonic with a fixed choice of $k$.  This is because
according to (\ref{HarmScaling}), the second-order differential
equation (\ref{ZDiffEq}) has only one solution regular at $\psi = 0$.
This uniqueness implies that for given $k$, there is no more than one
$SU(2)_H \times SU(2)_V$ representation labeled by $(\ell, \ell')$.

We can furthermore show that only $SU(2)_H \times SU(2)_V$
representations with $\ell + \ell' \leq k$ will appear inside the
$SO(6)$ representation labeled by $k$.  Recall that the $SO(6)$
representation is the $k$-fold symmetric product of the fundamental
{\bf 6}.  This decomposes into representations $({\bf 3}, {\bf 1})
\oplus ({\bf 1}, {\bf 3})$, {\i.e.} into a sum of $(\ell, \ell') =
(1,0)$ and $(\ell, \ell') = (0,1)$.  We easily see that the $k$-fold
product of this sum contains only representations satisfying $\ell +
\ell' \leq k$, with equality only when the factors in each $SU(2)$ are
completely symmetrized.

\section*{Appendix B: Field Theory Conventions}

We work in mostly-minus signature.  Minimal three-dimensional spinors
are Majorana, so it is convenient for us to use Majorana notation in
four dimensions as well.  A convenient Majorana basis for 3D ($2
\times 2$) and 4D ($4 \times 4$) Clifford matrices $\rho^k$ and
$\gamma^\mu$ is
\begin{eqnarray}
\label{3DGamma}
\rho^0 &=& - \sigma^2 \,, \quad \quad \rho^1 = i \sigma^1, \quad \quad
\rho^2 = i \sigma^3 \,, \\
\label{4DGamma}
\gamma^0 &=& \rho^0 \otimes \sigma^3 \,, \quad \quad
\gamma^1 = \rho^1 \otimes \sigma^3 \,, \quad \quad
\gamma^2 = \rho^2 \otimes \sigma^3 \,, \quad \quad
\gamma^3 = I \otimes i \sigma^1 \,, \quad \quad \,.
\end{eqnarray}
with $\sigma^k$ the Pauli matrices.  These matrices are all imaginary,
and $\rho^0$ and $\gamma^0$ are hermitian while the rest are
antihermitian.  In this basis, Majorana spinors are real in both three
and four dimensions.  We define the 4D chirality and projection
matrices as
\begin{eqnarray}
\label{gamma}
\gamma \equiv -i \gamma^0 \gamma^1 \gamma^2 \gamma^3 = I \otimes \sigma^2 \,,
\quad \quad L \equiv \tf12 (1 + \gamma) \,, \quad 
R \equiv \tf12 (1 - \gamma) \,,
\end{eqnarray}
with $\gamma$ pure imaginary and hermitian and as usual satisfying
$\gamma^2 = 1$, $\{ \gamma , \gamma^\mu \} = 0$.

%--------+---------+---------+---------+---------+---------+---------+
%Bibliography

\begingroup\raggedright\endgroup

\end{fmffile}
\end{document}